\documentclass[fleqn,usenatbib]{mnras}

\usepackage{newtxtext,newtxmath}
\usepackage{orcidlink}

\usepackage[T1]{fontenc}
\usepackage{caption}
\usepackage{subcaption}
\usepackage{ulem}
\DeclareRobustCommand{\VAN}[3]{#2}
\let\VANthebibliography\thebibliography
\def\thebibliography{\DeclareRobustCommand{\VAN}[3]{##3}\VANthebibliography}

\usepackage{graphicx}	
\usepackage{amsmath}	
\usepackage{xspace}

\title[Same galaxy, different CGM]{Same galaxy, different CGM: how the metal loading of galactic winds regulates the baryon cycle in Milky Way-mass galaxies}

\author[P. Khatri et al.]{Prachi Khatri$^{1}$\orcidlink{0009-0009-1983-8333},\thanks{E-mail: khatrip@cardiff.ac.uk}
Freeke van de Voort$^{1}$\orcidlink{0000-0002-6301-638X},
Rebekka Bieri$^{2}$\orcidlink{0000-0002-4554-4488},
Rüdiger Pakmor$^{3}$\orcidlink{0000-0003-3308-2420},
Robert J. J. Grand$^{4}$\orcidlink{0000-0001-9667-1340},  
\newauthor
Thomas A. Rintoul$^{1}$\orcidlink{0009-0006-9774-0807},
Maria Werhahn$^{3}$\orcidlink{0000-0003-4984-4389},
and Rosie Y. Talbot$^{3}$\orcidlink{0000-0001-9393-7879}
\\
$^{1}$Cardiff Hub for Astrophysics Research and Technology, Department of Physics and Astronomy, Cardiff University, Queen’s Buildings, Cardiff CF24 3AA, UK\\
$^{2}$Institut für Astrophysik, Universität Zürich, Winterthurerstrasse 190, 8057 Zürich, Switzerland\\
$^{3}$Max-Planck-Institut für Astrophysik, Karl-Schwarzschild-Str. 1, D-85748, Garching, Germany\\
$^{4}$Astrophysics Research Institute, Liverpool John Moores University, 146 Brownlow Hill, Liverpool, L3 5RF, UK
}

\date{Accepted XXX. Received YYY; in original form ZZZ}

\pubyear{\the\year{}}

\begin{document}
\label{firstpage}
\pagerange{\pageref{firstpage}--\pageref{lastpage}}
\maketitle

\newcommand{\elf}{$\epsilon_{w}$\xspace}
\newcommand{\mlf}{$\tilde{\eta}_{\mathrm{metal}}$\xspace}
\newcommand{\msun}{$\mathrm{M_{\odot}}$\xspace}
\newcommand{\zsun}{$\mathrm{Z_{\odot}}$\xspace}
\newcommand{\Zgas}{$Z_{\mathrm{gas}}$\xspace}
\newcommand{\Zstar}{$Z_*$\xspace}
\newcommand{\red}[1]{\textcolor{red}{#1}}
\newcommand{\blue}[1]{\textcolor{cyan}{#1}}

\begin{abstract}
The circumgalactic medium (CGM) is both the reservoir of gas that fuels star formation and galaxy growth, and the repository for the mass, energy, and metals expelled through stellar evolution and feedback. We present a controlled experiment using a suite of five cosmological magnetohydrodynamical simulations of a Milky Way-mass halo from the Auriga suite, wherein we vary the metal content and energy loading of galactic winds driven by stellar feedback. We chose these parameters in combination such that all five runs obtain very similar (within $\approx$10\%) $z=0$ stellar masses and morphologies and thus produce similar amounts of metals throughout their lifetimes. Our simulations differ substantially in where these metals end up, spatially within the halo and across the different baryonic components (stars, gas disc, CGM, and the intergalactic medium). The metal content affects the cooling efficiency of the halo gas, regulating the CGM cool gas fraction and the accretion rates onto the galaxy. Between our two extreme models, we report an order of magnitude difference in the median gas metallicity at vertical heights $|z| \gtrsim 30$ kpc above the disc plane. Current constraints on the metallicity of the Milky Way halo gas from absorption-line measurements of intermediate- and high-velocity clouds at $|z| \lesssim 15$ kpc are broadly consistent with our simulations, though a direct comparison is limited by differences in the adopted methods. Our findings show that galaxies with nearly identical stellar content and morphology can differ substantially in their CGM, which offers a promising avenue for constraining feedback physics within galaxy formation models.

\end{abstract}




\section{Introduction}
Galactic feedback is essential to produce realistic galaxies in cosmological simulations \citep{schaye2010, dave2011a, puchwein2013, vogelsberger2013}.
Because of the wide range of spatial and temporal scales involved in modelling feedback alongside other physics relevant for galaxy formation in a cosmological context, it is not feasible to model these processes from first principles. Instead, all galaxy formation models use effective (subgrid) feedback prescriptions that approximate the physical and chemical processes taking place on scales unresolved in cosmological simulations. Their role is to capture the impact of those processes on the scales we \textit{do} resolve, and thereby make testable predictions \citep[e.g.][]{vogelsberger2020}. 

Such prescriptions have various parameters that control the strength of the feedback and their values are usually chosen to match a set of empirical constraints. Many of these constraints relate to the stellar content alone, e.g., the mass of the galaxy produced in a Milky Way-mass halo and its stellar morphology, while others compare the statistical properties of the simulated galaxy population against observations, e.g., the stellar mass function \citep{schaye2015, pillepich2018a, dave2019, kugekl2023, chaikin2026}.

However, galaxies do not evolve in isolation. They are part of a wider ecosystem comprising the \textit{galaxy}, the gaseous halo surrounding it, commonly known as the \textit{circumgalactic medium (CGM)}, and the gas beyond the galaxy's dark matter halo that fills the large-scale structure of the cosmic web, the \textit{intergalactic medium (IGM)}. The IGM supplies the halo with fresh gas to sustain star formation over many Gyr \citep{keres2005, vandeVoort2011, grand2019}. Heavy elements (metals) are synthesised in stars and returned to the interstellar medium (ISM) through stellar evolution and feedback. Outflows launched by feedback (including stellar feedback) carry this metal-rich gas beyond the ISM where it mixes with the CGM \citep{tumlinson2011}. Winds driven by stellar feedback are ubiquitous in star-forming galaxies and play a crucial role in regulating galaxy growth and the properties of the CGM \citep{veilleux2020}. The metals they deposit both aid the cooling of halo gas and allow us to observe the CGM through emission and absorption lines of various metal ions.
Depending on the halo mass and the strength of the feedback, some of the metal-rich winds enrich the IGM as well. The previously outflowing gas can also be re-accreted onto the galaxy via fountain flows \citep[see e.g.,][]{marinacci2010, grand2019}. Thus, an effective baryon cycle operates among the components of this galactic ecosystem \citep[see][for a review]{peroux_howk_2020}.  The CGM and IGM properties can therefore provide meaningful constraints on the physical processes taking place within galaxies \citep[see][for a review]{tumlinson2017, crain2023, faucher-giguere2023}.

Despite this, the CGM is rarely used to calibrate galaxy formation models, partly because the available observational constraints are sparse. Therefore, it can serve as an independent test of feedback physics. If more than one set of feedback parameters can produce the ``same'' galaxy in terms of its stellar content and structure, then stellar constraints alone cannot distinguish between them. In this work, we devise a controlled experiment in which different sets of feedback parameters produce galaxies with nearly identical stellar content. We then examine if and how much their CGM properties differ.


For over a decade, the CGM has been advocated as a discriminator of the feedback models employed in cosmological simulations of galaxy formation \citep{hummels2013, shen2013, suresh2015, oppenheimer2018, ji2020, smith2024, bennett2025, rey2025, rey2026}. However, these studies typically compare different types of feedback models in a single galaxy or across a statistical sample of galaxies. 
Here we instead take a single, effective prescription that models galactic winds driven by stellar feedback and vary the energy and metal content of the winds, while holding the final stellar mass nearly fixed, to investigate how these changes impact the CGM in our simulated galaxies.

Towards this goal, we perform simulations of a Milky Way-mass halo, varying the metal content and the energy carried by the winds. These wind parameters are chosen such that the final stellar mass in the halo at $z=0$ lies within $\approx$10\% of the fiducial value. This spread is comparable to the intrinsic 
variability of the underlying galaxy formation model \citep[see][for more details]{pakmor2025a}. Each of our feedback models is thus an equally plausible way to produce the stellar disc of a Milky Way-like galaxy. We find that these galaxies, while nearly identical in their stellar content, differ substantially in the distribution of metals throughout their CGM.
Therefore, the CGM retains an imprint of feedback that the stellar component does not, opening a promising avenue for constraining feedback physics in future studies.

The paper is organised as follows. In Section~\ref{sec:methods}, we briefly describe the galaxy formation model and introduce our simulation suite. Section~\ref{sec:results} presents our results on the metal distribution within the halo, the phase structure of the CGM, and the inflow and outflow rates. We present a discussion of our findings including a comparison to previous work in Section~\ref{sec:discussion}. We summarise our results 
in Section~\ref{sec:conclusions}.

\section{Simulations and Methods}
\label{sec:methods}

\begin{table*}
    \caption{Parameters of the wind model for different runs in our simulation suite. The energy loading, mass loading, and wind specific energy are expressed with respect to their fiducial values.}
    \centering
    \begin{tabular}{c|c|c|c|c|c}
        & Metal loading & Energy loading & Mass loading & Wind velocity & Specific energy \\
        & \mlf & \elf & $\eta_w$ & $v_w$ & \elf / $\eta_w$ \\
        \hline
        \hline
        & 1.00 & 1.0 & 1.0 & same as fiducial & same as fiducial \\
        Fiducial & 0.40 & 1.0 & 1.0 & 3.46 $\sigma_{\mathrm{DM}}$ & 1.0\\
        & 0.25 & 0.8 & 0.8 & same as fiducial & same as fiducial \\
        & 0.20 & 0.7 & 0.7 & same as fiducial & same as fiducial \\
        & 0.10 & 0.5 & 0.5 & same as fiducial & same as fiducial \\
        \hline
    \end{tabular}
    \label{tab:runs}    
\end{table*}
\begin{table*}
    \caption{Definitions of the different gas components. The projected radial distance $r_{\mathrm{2D}}$ is the 2D radial distance within the disc defined such that the normal to the disc plane is along the angular momentum direction of all stars within the halo.}
    \centering
    \begin{tabular}{c|c|c|c|c|c|c|c}
        Component & $r_{\mathrm{3D}}$ range & $r_{\mathrm{2D}}$ range & $|z|$ range & Any other threshold\\
        \hline
        \hline
        Halo & $\le R_{200c}$ & - & - & satellites excluded \\
        Disc & $\le R_{200c}$ & $\le 50$ kpc & $\le 10$ kpc & satellites excluded\\
        CGM  & $\le R_{200c}$ & $> 50$ kpc & $> 10$ kpc & SFR=0; satellites excluded\\
        IGM & $> R_{200c}$& - & - & - \\
        \hline
    \end{tabular}
    \label{tab:definitions}    
\end{table*}
\begin{table*}
    \caption{Global properties of the galaxy and halo in our simulations at $z=0$. From left to right, the columns denote the metal loading \mlf; the energy loading \elf; the total stellar mass, $M_*$, within $R_{200c}$; the mean stellar metallicity, $\langle Z_*^{\mathrm{halo}} \rangle$, and age, $\langle \mathrm{age_*^{\mathrm{halo}}} \rangle$, of stars within $R_{200c}$; the mean age, $\langle \mathrm{age}_{*}^{\mathrm{5 \, kpc}}\rangle$, of stars within 5 kpc of the galactic centre and their total mass, $M_*^{\mathrm{5 \, kpc}}$, and the mean metallicity, $\langle Z_*^{\mathrm{5 \, kpc}} \rangle$, of these stars; the scale radius, $r_s$, and scale height, $h_s$, of the stellar disc; the total gas mass, $M_{\mathrm{gas}}^{\mathrm{halo}}$, within $R_{200c}$; the total gas mass, $M_{\mathrm{gas}}^{\mathrm{disc}}$, within the disc (as defined in Table~\ref{tab:definitions}) and the mean metallicity, $\langle Z_{\mathrm{gas}}^{\mathrm{disc}} \rangle$, of this gas; the total mass, $M_{\mathrm{gas}}^{\mathrm{CGM}}$, and the mean metallicity, $\langle Z_{\mathrm{gas}}^{\mathrm{CGM}} \rangle$, of the CGM; the total mass of metals, $M^{\mathrm{metal}}_{\mathrm{total}}$, produced in all stars within the halo (excluding satellites) until the present day; and the fraction of these metals present in different baryonic components, $f^{\mathrm{metal}}_{i}$. 
    All mean values are computed weighted by the mass of the relevant star particles or gas cells. The run with \mlf = 0.4 and \elf = 1.0 corresponds to the fiducial set of parameters in the Auriga model.}
    \centering
    \setlength{\tabcolsep}{0.5pt}
    \resizebox{\textwidth}{!}{
    \begin{tabular}{c|c|c|c|c|c|c|c|c|c|c|c|c|c|c|c|c|c|c|c|c}
         \mlf & \elf & $M_*$ & $\langle Z_*^{\mathrm{halo}}\rangle $& $\langle \mathrm{age}_*^{\mathrm{halo}}\rangle $ &
         $\langle \mathrm{age}_{*}^{\mathrm{5 \, kpc}}\rangle $ & $M_*^{\mathrm{5 \, kpc}}$ & $\langle Z_*^{\mathrm{5 \, kpc}}\rangle $ &  $r_s$ & $h_s$ &
         $M^{\mathrm{halo}}_{\mathrm{gas}}$ & $M^{\mathrm{disc}}_{\mathrm{gas}}$ & $\langle Z^{\mathrm{disc}}_{\mathrm{gas}}\rangle $&  
         $M^{\mathrm{CGM}}_{\mathrm{gas}}$ & 
         $\langle Z^{\mathrm{CGM}}_{\mathrm{gas}}\rangle $  & $M^{\mathrm{metal}}_{\mathrm{total}}$& $f^{\mathrm{metal}}_*$ & $f^{\mathrm{metal}}_{\mathrm{disc}}$ & $f^{\mathrm{metal}}_{\mathrm{CGM}}$ & $f^{\mathrm{metal}}_{\mathrm{IGM}}$  \\
         & & [$10^{10}$ \msun]   & [\zsun] & Gyr &  Gyr & [$10^{10}$ \msun] & [\zsun] & [kpc] & [kpc] & [$10^{10}$ \msun]   & [$10^{10}$ \msun]   & [\zsun] &  
         [$10^{10}$ \msun] & [\zsun]  & [$10^{9}$ \msun] &
         \\
         \hline
         \hline
         1.00 & 1.0 & 6.09 & 1.46 & 7.00 & 5.58 & 1.89 & 1.40 & 5.47 $\pm$ 0.07 & 1.28 $\pm$ 0.06& 
         6.25 &4.22&2.19&
         1.68 & 1.04 & 3.38 & 0.33 & 0.35 & 0.09 & 0.22 \\
         0.40 & 1.0 & 5.68 & 1.99 & 7.31 & 5.94 & 1.64 & 2.74& 4.68 $\pm$ 0.11 & 1.27 $\pm$ 0.05& 
         6.14 & 3.57&1.98&
         2.02 & 0.60 & 3.12  & 0.46 & 0.29 & 0.08 & 0.18 \\
         0.25 & 0.8 & 5.6 & 2.53 & 6.51  & 5.92 & 2.50 & 3.57& 3.84 $\pm$ 0.05 & 1.08 $\pm$ 0.07& 
         6.43 & 3.54&2.02&
         2.21&0.32& 3.10 &0.58 & 0.29 & 0.07 & 0.05 \\
         0.20 & 0.7 & 5.6 & 2.59 & 6.83 & 5.99 & 2.25 & 3.83 & 4.05 $\pm$ 0.05& 1.12 $\pm$ 0.08& 
         6.51 & 3.46&1.93&
         2.52&0.27& 3.06 &0.60 & 0.28 & 0.06 & 0.05 \\
         0.10 & 0.5 & 6.08 & 2.84 & 6.99 & 5.96  &2.52 & 4.35 & 3.82 $\pm$ 0.10& 1.06 $\pm$ 0.08&
         6.44 & 3.22&2.17&
         2.84&0.20& 3.31 & 0.66 & 0.27 & 0.03 & 0.03 \\
         \hline
    \end{tabular}}
    \label{tab:global}
\end{table*}

\begin{figure*}
	\includegraphics[width=\textwidth, trim={0 2cm 0 0}, clip]{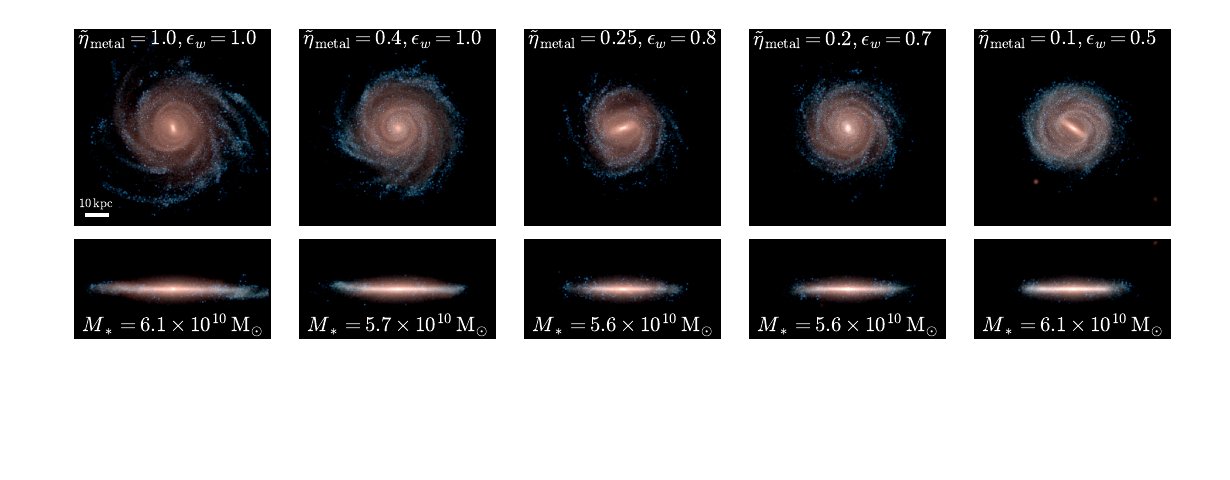}
    \caption{Stellar light projections of the galaxies in the five runs with the parameters \mlf and \elf listed in the top panel. All projections span a line-of-sight distance of 80 kpc. The face-on projection (top) are shown for a 80 kpc $\times$ 80 kpc box and the edge-on projections (bottom) for a 80 kpc $\times$ 40 kpc box. The images show the $K$-band, $B$-band, and $U$-band luminosity of stars using the red, green, and blue colour channels, respectively, in logarithmic intervals. Younger (older) star particles are therefore represented by bluer (redder) colours. All five galaxies have similar stellar masses (indicated in the bottom panels) and look similar in their stellar distribution. The lower \mlf runs on the right feature more compact discs compared to the higher \mlf runs on the left.}
    \label{fig:stellar_maps}
\end{figure*}

In this study, we perform cosmological magnetohydrodynamical zoom-in simulations of a Milky Way-mass halo from the Auriga simulation suite \citep{grand2017, grand2024}, using the full Auriga galaxy formation model. We perform five simulations of the same dark matter halo (Au-6), varying two parameters of the wind feedback prescription in each simulation.
All simulations begin at $z=127$ with cosmological parameters $\Omega_{\mathrm{m}} = 0.307$, $\Omega_{\mathrm{b}} = 0.048$, $\Omega_{\Lambda} = 0.693$, and Hubble constant $H_0 = 100 \, h \, \mathrm{km \, s^{-1} \, Mpc^{-1}}$, with $h=0.6777$ \citep{planck2014}. 
We use the moving-mesh magnetohydrodynamic code \textsc{arepo} \citep{arepo, pakmor2016, weinberger2020}. \textsc{arepo} solves the magnetohydrodynamical equations on an unstructured Voronoi mesh that moves with the fluid flow, resulting in a quasi-Lagrangian treatment of the fluid equations. The dark matter particle mass in our simulations is $\sim 4 \times 10^5$ \msun. The gas resolution is set by the target gas mass $m_{\mathrm{target}} \sim 5 \times 10^4$ \msun. Because the mesh in \textsc{arepo} is only quasi-Lagrangian, mass is still exchanged between the cells. Therefore, the cells are explicitly refined or derefined to keep their masses within a factor of two of the target gas mass. The galaxy formation model is described in full detail in \cite{grand2017}. Here we summarize some key aspects of the model, with a particular focus on the ones we vary in this study.

The galaxy formation model includes primordial and metal-line cooling with self-shielding following \cite{rahmati2013}, and a spatially uniform UV background from reionization completed at $z=6$ \citep{faucher-giguere2009, vogelsberger2013}, magnetic fields \citep{pakmor2013, pakmor2014, pakmor2017}, gas accretion onto black holes, and feedback from active galactic nuclei.

Star formation is modelled following the subgrid model of \cite{springel-hernquist2003}, in which gas cells above a density threshold corresponding to a hydrogen number density $n_{\mathrm{H}} =  0.11 \mathrm{cm}^{-3}$ are treated as a two-phase medium of cold clouds in pressure equilibrium with a hot phase. Because the simulations do not resolve this multiphase structure of the ISM, this prescription sets an effective equation of state for star-forming gas and provides a density-dependent star formation rate (SFR). In turn the SFR determines the probability of stochastically forming star particles. 

Each star particle represents a single stellar population (SSP) of a given age and metallicity. The distribution of stellar masses in each SSP is given by the \cite{chabrier2003} initial mass function (IMF). The star particle inherits the metallicity of its parent gas cell at birth. The IMF allows us to calculate, at each time step, the mass of the SSP that moves off the main sequence. Based on this we can calculate the mass fraction that enters the asymptotic giant branch (AGB) phase, or explodes as supernova Type Ia (SNIa) or a core-collapse supernova (Type II supernova, SNII). The number of SNII in a given SSP is calculated from the number of stars within this SSP in the mass range 8-100 \msun. The mass and metal return fractions of each of these stellar feedback events is calculated based on yield tables. For AGB stars, we use the yields from \cite{karakas2010} and for SNII the yields from \cite{portinari1998}.  The rate of SNIa events is computed using a delay time distribution. The amount of mass and metals returned to the ISM for each SNIa event is calculated based on yield tables from \cite{thielemann2003} and \cite{travaglio2004}. The mass and metals released are deposited in the host cell only, rather than distributing these to 64 nearest neighbours as done in the original Auriga simulations \citep{grand2017}.

Galactic winds generated by stellar feedback are modelled using an effective prescription, which specifies the wind properties such as its energy, mass, and metal content, but is agnostic to the specific source of the wind. Following \cite{vogelsberger2013}, the occurrence of winds is decided by the dimensionless mass loading parameter $\eta_w = \dot{M}_w / \dot{M}_{\mathrm{SFR}}$, which sets the wind mass flow rate relative to the star formation rate. Because wind launching is probabilistic, the ratio holds in a statistical manner. Practically, this is implemented as follows: at a given time step, $\Delta t$, the $i$-th gas cell (above the star formation density threshold) is selected to participate in star formation or wind generation with a probability 
\begin{equation}
    p = \frac{M_i}{M_*} \, \left( 1 - e^{-(1+\eta_w) \Delta t / t_{\mathrm{SF}}} \right),
\end{equation}
where $M_i$ is the mass of the gas cell under consideration, $M_*$ is the star and wind particle mass, and $t_{\mathrm{SF}}$ is the star formation timescale. If a given gas cell is selected, 
then a number is drawn from the uniform random distribution $x \in U(0,1)$; the cell forms a star particle if $x < 1/(1+\eta_w)$, otherwise, it gets ejected as a wind particle.

Thus, a selected gas cell forms a star particle or gets ejected as a wind particle with the following probabilities:
\begin{equation}
\label{eq:p_sf_wind2}
    p_{\mathrm{SF}, w} =
    \begin{cases}
        1/(1+\eta_w)      & \text{for star formation}, \\
        \eta_w/(1+\eta_w) & \text{for winds}.
    \end{cases}
\end{equation}

The wind particle is launched in a direction randomly chosen from an isotropic distribution. The ejection speed of the wind particle ($v_w$) is set by the local one-dimensional dark matter velocity dispersion $\sigma_{\mathrm{DM}}$ \citep{okamoto2010}, calculated from the 64 nearest dark matter particles as: 
\begin{equation}
    \label{eq:wind_velocity}
    v_w = \kappa_{\mathrm{kin}} \, \sigma_{\mathrm{DM}} \, ,
\end{equation}
where $ \kappa_{\mathrm{kin}}$ is a dimensionless effective model parameter. We set $\kappa_{\mathrm{kin}} = 3.46$. This choice is broadly consistent with \cite{okamoto2010}, who showed that such a scaling (with proportionality factors $\sim$4-5) reproduces the observed satellite luminosity function and the luminosity-metallicity relation of Local Group satellites in their simulations. More generally, such a wind parametrization is required to reproduce the stellar mass and oxygen abundances of low-mass haloes \citep{grand2019, puchwein2013}.

The total energy, $E_w$, available to drive the wind is set by the dimensionless wind energy loading factor, \elf, expressed in units of the canonical SNII energy of $10^{51}$ erg. The energy available to drive the wind per unit stellar mass formed is then equal to
\begin{equation}
    E_w = \epsilon_w \, N_{\mathrm{SNII}} \, 10^{51} \, \mathrm{erg} \, ,
\end{equation}
where $N_{\mathrm{SNII}}$ is the number of SNII events per unit stellar mass formed. For our adopted IMF and stellar evolution model, $N_{\mathrm{SNII}} = 1.73 \times 10^{-2} \, \mathrm{M_{\odot}^{-1}}$. We note that the SNII energy per unit stellar mass formed is only used here as a convenient reference and the wind itself could be generated by sources other than SNII, such as stellar winds, stellar radiation, and cosmic rays. 

The wind energy is related to the wind velocity and mass loading as:
\begin{equation}
    \label{eq:mass_loading}
    \begin{aligned}
        E_w 
        &= \frac{1}{2} \, \eta_w \, v_w^2 \,+\, \frac{3}{2} \, \eta_w \, \kappa_{\mathrm{th}} \, \sigma^{2}_{\mathrm{DM}} \\
        &= \frac{1}{2} \, \eta_w \, \sigma^{2}_{\mathrm{DM}} \, \left( \kappa_{\mathrm{kin}}^2 \,+\, 3 \, \kappa_{\mathrm{th}} \right) \, .
    \end{aligned}
\end{equation}
The parameters $\kappa_{\mathrm{kin}}$ and $\kappa_{\mathrm{th}}$, respectively, set the velocity (Equation~\ref{eq:wind_velocity}) and temperature of the wind. In the Auriga model, $\kappa_{\mathrm{th}}=3$. Similar to $\kappa_{\mathrm{kin}}$, $\kappa_{\mathrm{th}}$ is also a dimensionless effective model parameter.

The wind particle has a metallicity equal to \mlf times the metallicity of the gas cell from where it was launched, where \mlf is called the wind metal loading parameter\footnote{We note that our 1- \mlf is equivalent to the $\eta_w$ parameter defined in \cite{grand2017}. Instead, here we use $\eta_w$ to represent the wind mass loading.}. The remaining metal mass is distributed among nearby cells. We restrict ourselves to metal-depleted winds, where the wind metallicity is always smaller than the midplane ISM metallicity. In other words, we assume that the winds are dominated not by direct supernova ejecta but by the entrained gas, a substantial fraction of which itself is more metal-poor compared to the midplane ISM. If instead the outflows were primarily direct supernova ejecta, the metallicity of the wind might be much higher than the ISM metallicity because the metals from the supernova ejecta would be transported directly into the outflow without first mixing with the surrounding ISM. We further discuss the plausibility of such metal-depleted winds in Section~\ref{sec:low_Z_wind}.

Upon launch, the wind particle decouples from the gas. Initially interacting only gravitationally, it recouples to the gas once it reaches an ambient density of 5 per cent of the star formation density threshold or after a certain maximum time has lapsed. On recoupling it deposits its mass, metals, energy, and momentum into the gas cell where it is present. 

In the fiducial Auriga model, the metal loading and energy loading are set to 0.4 and 1.0, respectively. In this work, we design a controlled experiment where different set of wind feedback parameters (\mlf and \elf) produce the same (within $\approx$10\%) final stellar mass within the halo. The allowed variation of 10\% was chosen based on a similar level of realisation-to-realisation scatter found in \cite{pakmor2025a}. In addition to the fiducial parameter set, we consider three lower values of the metal loading (\mlf = 0.25, 0.2, and 0.1). Reducing \mlf while keeping the energy loading fixed at its fiducial value changes the final stellar mass by more than our allowed tolerance of 10\%. We therefore adjust \elf for each value of \mlf such that the final stellar mass remains within 10\% of the fiducial run. Thus, the values of \mlf and \elf are not varied independently, but are chosen in combination to construct a sample with approximately fixed final stellar mass. 

We also consider a metal loading higher than the fiducial value, \mlf =  1.0. 
In this case, keeping the energy loading at its fiducial value of 1.0 results in a slightly higher final stellar mass, but the difference remains within our 10\% tolerance. We therefore include this parameter combination in our controlled sample. We also tested \elf = 1.3 in combination with \mlf = 1.0, but this produced a substantially lower final stellar mass and we therefore exclude this run.
The metal and energy loading of the five runs included in this work are listed in Table~\ref{tab:runs}, along with other model parameters for reference.

For all simulations, we define the virial radius, $R_{200c}$, as the radius enclosing a mean density of 200 times the critical density of the Universe at the given redshift. For each galaxy, the gas disc is defined such that the normal vector to the disc plane is aligned with the angular momentum direction of all stars within the halo. The disc region is defined to have a projected radius $r_{\mathrm{2D}} = 50$ kpc and a height of 10 kpc above and below the plane, where the 50 kpc radius is chosen to be large enough to enclose the neutral hydrogen dominated extended disc in every run (see Figure~\ref{fig:gas_maps}). We refer to all gas within the halo but outside the disc as the CGM. Throughout this paper, we adopt these definitions of the gas disc and the CGM and they are listed in Table~\ref{tab:definitions}.  


\section{Results}
\label{sec:results}

\begin{figure}
	\includegraphics[width=\columnwidth]{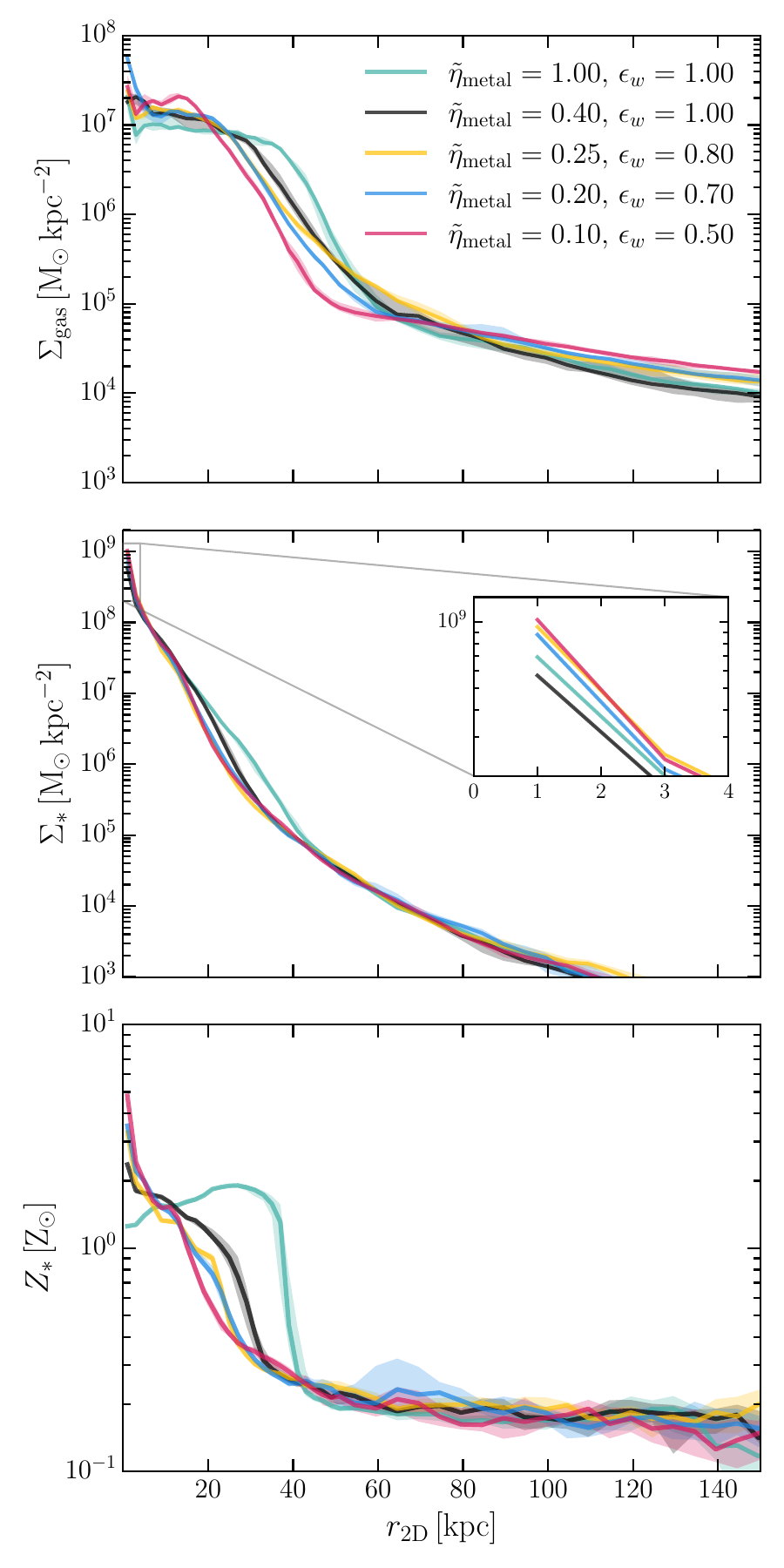}
    \caption{Gas mass surface density ($\Sigma_{\mathrm{gas}}$; top), stellar mass surface density ($\Sigma_{*}$;middle), and stellar metallicity (\Zstar; bottom) profiles as a function of the projected radial distance ($r_{\mathrm{2D}}$) in a face-on projection computed for gas within $|z|< 10$ kpc of the disc plane. The solid lines show the median of the per-snapshot mass-weighted medians in a given radial bin over the last 18 snapshots covering the final Gyr of evolution. The shaded bands around the profiles show the 16-84 percentile range of these medians and represent the temporal scatter. In all three panels, the profiles for different runs are broadly similar to each other at all radii, except for the \Zstar profile of the highest \mlf run. The gas mass surface density starts to decline at smaller radii for the lower \mlf runs as a result of the relatively smaller extended gas discs in these (also see Figure~\ref{fig:gas_maps}). Similarly, the $\Sigma_*$ profiles show minor differences as a result of the different disc sizes. The \Zstar profile for our highest \mlf run is strikingly different from the other runs and shows a positive gradient in the inner $\sim 20$ kpc. }
    \label{fig:stellar_surface_density}
\end{figure}
\begin{figure}
    \includegraphics[width=\columnwidth]{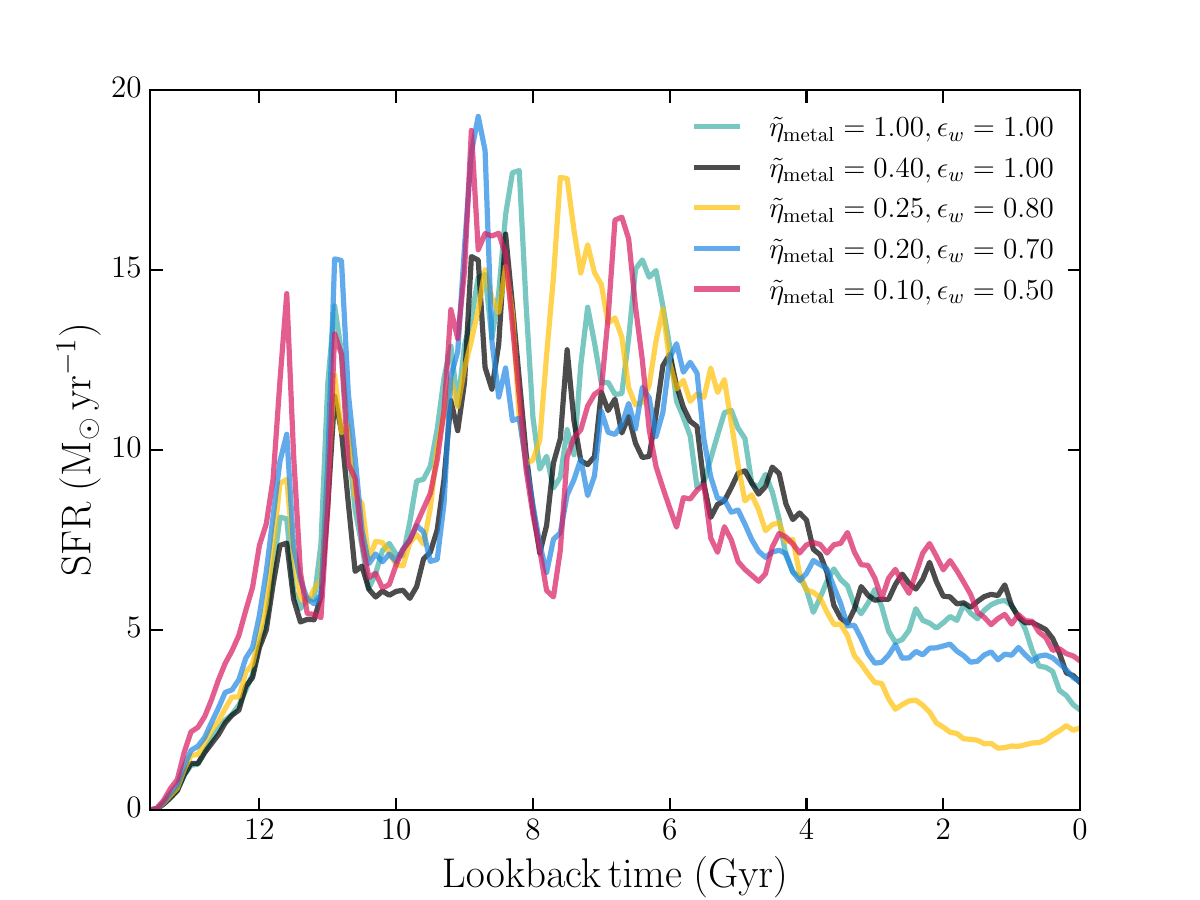}
    \caption{Star formation histories (averaged over a 100 Myr timescale) as a function of the lookback time computed from the initial mass and ages of all stars within the halo in each run. In the first couple of Gyr, the variations in the star formation histories are driven primarily by the different mass loading of winds (the higher \elf runs also have a higher mass loading; see Table~\ref{tab:runs}). At late times, variations are largely stochastic in nature. 
    }
    \label{fig:sfh}
\end{figure}

\subsection{Global galaxy properties}
\label{sec:stars}
We report in Table~\ref{tab:global} the total stellar mass ($M_*$) within the halo at $z=0$ (excluding satellites).
The $M_*$ values do not vary by more than 10\% across the runs. We note that a comparable spread was reported by \cite{pakmor2025a} for different realisations of Au-6 using identical model parameters (fiducial values for the wind parameters). These realisations differ only in the random number seed used for the stochastic sampling of star particle formation and wind particle ejection. Thus, the stellar masses of our runs do not vary more than the realisation-to-realisation scatter of the model and they are therefore indistinguishable based on stellar mass alone.

Figure~\ref{fig:stellar_maps} shows the stellar light projections of the halo in the five simulations at $z=0$. From left to right, both the metal loading parameter \mlf and the energy loading parameter \elf of winds decrease (except for the same \elf in the first two runs). The top and bottom rows show, respectively, the face-on and edge-on projections of the stellar light. 
Despite large differences in the wind parameters, we find a remarkable similarity in the morphology of the stellar distributions. 
Moreover, the mean stellar ages are very similar across the runs, both globally (6.5-7.3 Gyr, a spread of 0.05 dex), and within the central 5 kpc (5.6-6.0 Gyr). The spread in both cases is well below the typical uncertainty on stellar ages derived from fitting the spectral energy distribution of galaxies.

The galaxies visibly differ in their bar strengths, with no systematic dependence on \mlf or \elf. \cite{pakmor2025a} find a similar variation among different realisations of Au-6 using the same feedback model and attribute it to the gas disc of this halo being only marginally bar-unstable. As a result, small fluctuations due to the intrinsic stochasticity of the galaxy formation model are amplified into visibly different bar strengths. The variations we see across our runs are therefore not driven by the wind feedback parameters.

Figure~\ref{fig:stellar_surface_density} shows the gas mass surface density ($\Sigma_{\mathrm{gas}}$), the stellar mass surface density ($\Sigma_{*}$), and the mass-weighted median stellar metallicity (\Zstar) profiles as a function of the 2D radial distance in a face-on projection for the five simulations. For the surface density profiles, we stack the last eighteen snapshots covering the final Gyr of evolution and show the median of these snapshots in the figure as a solid line, while the shaded band denotes the 16-84 percentile of the surface density in a given radial bin across the snapshots. For the metallicity profile, we compute the mass-weighted median stellar metallicity, \Zstar, in each radial bin for a given snapshot and show the median of these per-snapshot medians in each radial bin as a solid line in the figure. The shaded band in this case denotes the 16-84 percentile of the per-snapshot medians, and represents the temporal scatter in the median. 

For the gas and stellar surface density, we find only marginal differences across the runs. The one exception is the extent of the gas disc, which increases towards larger \mlf, ranging from $\sim$20 kpc in the lowest \mlf run to $\sim$40 kpc in the highest \mlf run. We find that the three lower \mlf runs (\mlf $\leq 0.25$) form more compact stellar discs; their central ($r_{\mathrm{2D}} < 2$ kpc) stellar surface density reaches $\approx$$10^9 \, \mathrm{M_{\odot} \, kpc^{-2}}$ (as seen from the inset in Figure~\ref{fig:stellar_surface_density}) compared to $\approx$$6-7 \times 10^8 \, \mathrm{M_{\odot} \, kpc^{-2}}$ in the two highest \mlf runs, and their $\Sigma_*$ profiles decline more steeply beyond $\approx$15 kpc, consistent with their smaller gas discs. The same trend is apparent in the stellar masses within the central 5 kpc: the three lower \mlf runs have $M_*^{\mathrm{5 \, kpc}}\sim$$2.3-2.5 \times 10^{10}$ \msun compared to the $M_*^{\mathrm{5 \, kpc}}\sim1.6-1.9 \times 10^{10}$ \msun in the two higher \mlf runs. The scale radius ($r_s$) and height ($h_s$) of the stellar disc are computed by fitting an exponential curve to the radial and vertical stellar surface density profiles, respectively, and are listed in Table~\ref{tab:global}. While $r_s$ increases with \mlf, $h_s$ does not vary significantly across the runs.

In Table~\ref{tab:global}, we also report the mass-weighted mean metallicity, $\langle Z_*^{\mathrm{halo}}\rangle$, of the stars within the halo and of those within the central 5 kpc, $\langle Z_*^{\mathrm{5 \, kpc}}\rangle$. We find that $\langle Z_*^{\mathrm{halo}}\rangle$ varies systematically with \mlf: the lower \mlf runs (\mlf $\leq 0.25$) have $\approx$25-40 per cent higher $\langle Z_*^{\mathrm{halo}}\rangle$ than the fiducial run, while the highest \mlf run has $\approx$25 per cent lower $\langle Z_*^{\mathrm{halo}}\rangle$. In the higher \mlf runs, the wind removes more metals from the star-forming gas, and thus less are available to get incorporated into successive generations of stars. This depletion of metals from stars is particularly more pronounced in the central regions, as seen from the values of $\langle Z_*^{\mathrm{5 \, kpc}}\rangle$. This results in mildly positive gradient in \Zstar within the inner 20 kpc of the stellar disc as opposed to the negative gradient shown by all other runs. We further inspect the gas metallicity profiles of these runs in Section~\ref{sec:gas_metallicity}.

Figure~\ref{fig:sfh} shows the star formation history (SFH) of the five runs averaged over 100 Myr time bins computed from all stars within $R_{200c}$, excluding stars within satellites. In the first couple of Gyr, the star formation rates are partly determined by the mass loading of winds. The lower wind mass loading in the lower \elf runs (see Table~\ref{tab:runs}) allows the galaxies to retain more gas mass in the ISM, which boosts early star formation. We also see a similar concentration of peaks across our runs at lookback times of 8-10 Gyr. Beyond that, the peaks and troughs in the SFH do not follow any systematic trends with the wind parameters and any variations tend to be mostly stochastic. We have verified this with two additional realisations of the fiducial run and throughout the paper show results for the run which has a stellar mass between the other two at $z=0$. The late time SFH also does not show any systematic trend with the adopted \mlf or \elf values. 

\subsection{The gas content and metal distribution}
\label{sec:gas_metallicity}

\begin{figure*}
	\includegraphics[width=0.9\textwidth, trim={0 0 0 0}, clip]{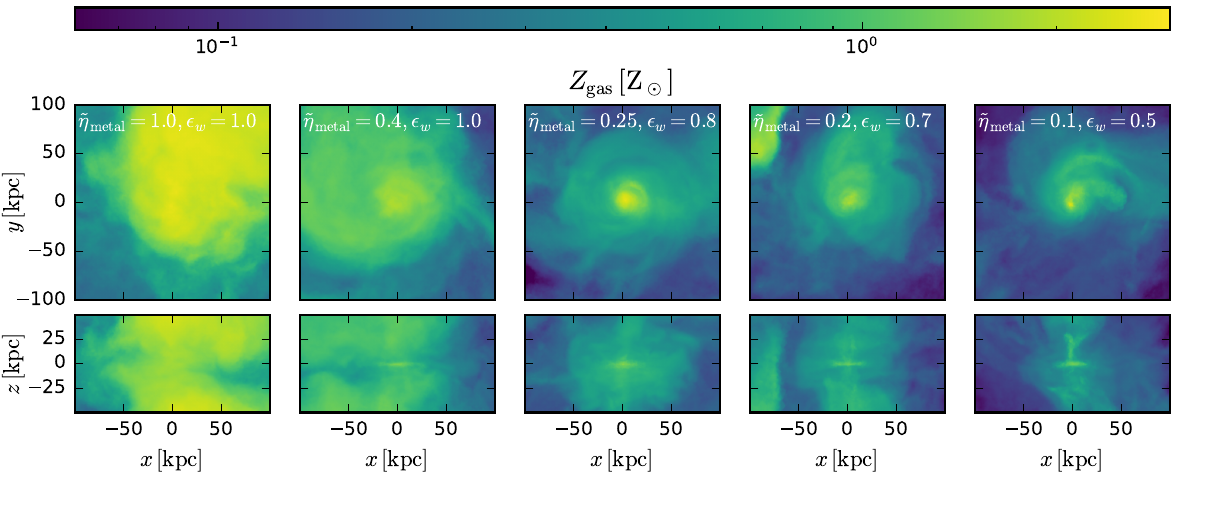}
    \caption{Face-on (top) and edge-on (bottom) projections of the gas metallicity across the different runs. For all projections, we compute the mass-weighted mean along a line-of-sight distance of 200 kpc. On the extreme left, our highest \mlf run shows gas enriched to solar and supersolar metallicities out to larger distances from the galactic centre. From left to right, the area covered by high metallicity gas progressively drops.}
    \label{fig:gas_metallicity_maps}
\end{figure*}
\begin{figure*}
	\includegraphics[width=0.95\textwidth,]{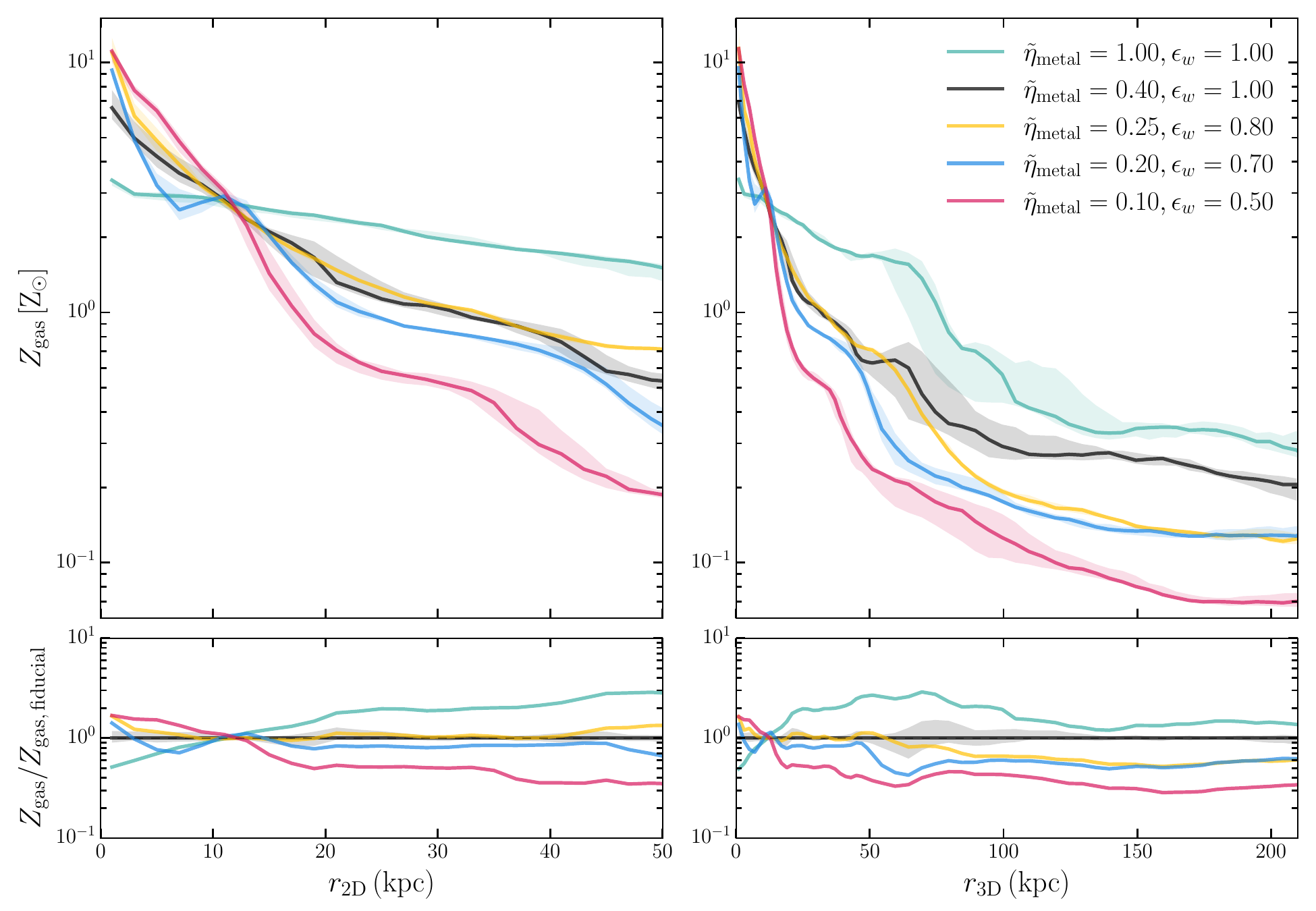}
    \caption{Gas metallicity (\Zgas) profiles as a function of the projected radial distance (from a face-on projection computed for gas within $|z|< 10$ kpc of the disc plane; left) out to 50 kpc and the galactocentric radius (right) out to $R_{200c}$ ($\sim 210$ kpc in all runs). Both profiles show the median over the last 18 snapshots covering the final 1 Gyr. For both panels, we use radial bins of size 2 kpc out to 50 kpc and size 5 kpc beyond that. As in Figure~\ref{fig:stellar_surface_density}, the solid line shows the median of the per-snapshot mass-weighted medians in a given radial bin and the shaded area represents the 16-84 percentile range of these medians. The bottom panels show the ratio of a given run and the fiducial run (shown in black). Compared to the fiducial run, the lower \mlf runs runs exhibit a higher central metallicity (up to a factor of 2) along with a steeper decline at large radii. On the other hand, the highest \mlf run shows a shallower decline out to $r_{\mathrm{3D}} \sim 60$ kpc. 
    }
    \label{fig:gas_metallicity_profile}
\end{figure*}
Similar to the extended stellar discs in Figure~\ref{fig:stellar_maps}, the higher \mlf runs also feature more extended gas discs (see Figure~\ref{fig:gas_maps}), while those for the lower \mlf runs are more compact (also see the $\Sigma_{\mathrm{gas}}$ profiles in Figure~\ref{fig:stellar_surface_density}). In Table~\ref{tab:global}, we report the total gas mass within the halo ($M^{\mathrm{halo}}_{\mathrm{gas}}$), the gas mass in the disc ($M^{\mathrm{disc}}_{\mathrm{gas}}$), and the gas mass in the CGM ($M^{\mathrm{CGM}}_{\mathrm{gas}}$).

All our runs have similar $M^{\mathrm{halo}}_{\mathrm{gas}}$ values (variations within 6 per cent), but this is split differently between the disc and the CGM. The runs with a higher energy loading and thus, a higher mass loading (following Table~\ref{tab:runs}), have a more massive extended gas disc and a less massive CGM. A higher mass loading in these runs means that by definition (Equation~\ref{eq:p_sf_wind2}), they launch a higher fraction of gas mass into winds per unit star formation and thus, naively, one would expect more gas mass in the CGM. We find results contrary to this expectation because in our simulation suite, the runs with a higher mass loading also have a higher metal loading, and so their CGM is more metal-rich (exact values reported in Table~\ref{tab:global}). The higher metal content of the CGM aids the cooling of gas, which then accretes onto the disc and builds up its mass over time. This also results in more gas cycling out of and back into the disc. In the Auriga model, such a fountain flow enhances the angular momentum of the gas \citep{grand2019}. This gas is deposited on the disc outskirts leading to a more extended gas disc over time. 

Figure~\ref{fig:gas_metallicity_maps} shows the face-on and edge-on projections of the mass-weighted mean gas metallicity along the line of sight in the five runs. Significant differences in the metallicity of the halo gas are apparent between the runs. Our \mlf = 1.0, \elf = 1.0 run contains gas enriched to solar and super-solar metallicities out to considerably larger distances in the halo compared to the other runs. For instance, its edge-on projection shows higher metallicities along the minor axis, extending to $\gtrsim 50$ kpc above the disc plane. For the same run, the metallicity along the major axis is about an order of magnitude lower. This azimuthal variation is driven by the outflows that escape along the path of least resistance perpendicular to the disc and predominantly deposit metals along the polar direction, as also seen in other simulations \citep{peroux2020}. A similar pattern is seen in the other runs, although both the opening angle and the radial extent of the metal-rich component progressively decrease. This trend is most pronounced in our \mlf = 0.1, \elf = 0.5 run, which reaches significantly lower metallicities above the disc plane and any higher metallicity gas outside the gas disc is confined to a narrow opening angle along the polar direction.

For gas in the disc and CGM, we also report, in Table~\ref{tab:global}, the mass-weighted mean metallicity, $\langle  Z^{\mathrm{disc}}_{\mathrm{gas}} \rangle$ and $\langle  Z^{\mathrm{CGM}}_{\mathrm{gas}} \rangle$, respectively. The fiducial run (with \mlf = 0.4, \elf = 1.0) has $\langle  Z^{\mathrm{CGM}}_{\mathrm{gas}} \rangle =0.6$ \zsun. Increasing \mlf to unity while keeping \elf fixed increases this to $\sim$\zsun. Since the wind energy loading is unchanged between these two runs, the increased halo metallicity is solely driven by the increased metal content of the winds, rather than a stronger outflow. Likewise, reducing \mlf below the fiducial value, together with a reduced \elf, decreases the $\langle  Z^{\mathrm{CGM}}_{\mathrm{gas}} \rangle$ substantially; our most extreme run on this end (with \mlf = 0.1, \elf= 0.5) has a three times lower mean CGM metallicity ($\langle  Z^{\mathrm{CGM}}_{\mathrm{gas}} \rangle =0.2$ \zsun) compared to the fiducial run. 

In Figure~\ref{fig:gas_metallicity_profile}, we focus on the gas disc and show the radial gas metallicity (\Zgas) profiles as a function of the 2D projected radius ($r_{\mathrm{2D}}$) and the 3D galactocentric radius ($r_{\mathrm{3D}}$). We find differences in the metallicity gradients across the runs both within the inner disc and at larger $r_{\mathrm{2D}}$. The highest \mlf run shows only a shallow decline in \Zgas across the disc, dropping by a factor of $\sim$2 out to $r_{\mathrm{2D}}$ = 50 kpc. A single powerlaw with slope $\approx$$-0.18$ describes the full profile, although it is essentially flat between 2 kpc $< r_{\mathrm{2D}}<$ 20 kpc, and the decline occurs at larger radii. The stellar metallicity profile for this run shows a mildly positive gradient in the inner 20 kpc (Figure~\ref{fig:stellar_surface_density}). This is because the stellar metallicities are derived from the  gas metallicities at the time each star forms, and not the present-day distribution.

In contrast, all the other runs (with \mlf $\leq 0.4$) show a much steeper decline out to  $r_{\mathrm{2D}} \sim 20$ kpc beyond which the decline is more gradual. The \Zgas profiles in these runs are best described by two separate powerlaws in the inner and outer disc. These runs also show a higher \Zgas in the central few kpc compared to the highest \mlf run. Overall, at $r_{\mathrm{2D}} = 50$ kpc, the lowest \mlf run has a factor of three lower \Zgas than the fiducial run, while the highest \mlf run has a factor of three higher \Zgas.

The right panel of Figure~\ref{fig:gas_metallicity_profile} shows the 3D radial profile. Here again we see that all profiles show a high metallicity towards the galactic centre with values of 3 \zsun for \mlf= 1.0, 5 \zsun for the fiducial \mlf= 0.4, and 10 \zsun for \mlf= 0.1. All profiles decline as we move outward, with the lower \mlf runs featuring a steeper decline. In the halo outskirts, the median metallicity in the \mlf = 1.0 run is $\approx$0.3 \zsun, in the fiducial run, it is 0.2 \zsun, and for the \mlf= 0.1 run it is $\approx$0.06 \zsun. We note that these values are different from the mean mass-weighted mean CGM metallicity ($\langle Z^{\mathrm{CGM}}_{\mathrm{gas}} \rangle$) values reported in Table~\ref{tab:global} as those are dominated by the inner CGM, which contains a large fraction of the mass.

In summary, Figures~\ref{fig:gas_metallicity_maps} and \ref{fig:gas_metallicity_profile} show that reducing the metal loading of galactic winds strongly affects the distribution of metals in both the gas disc and the CGM. We inspect the impact of these differences on the phase structure of the CGM in Section~\ref{sec:cgm_phase_structure} and further discuss the metal distribution in Section~\ref{sec:implications}.

\begin{figure}
	\includegraphics[width=0.5\textwidth, trim={1.2cm 0 0 0}, clip]{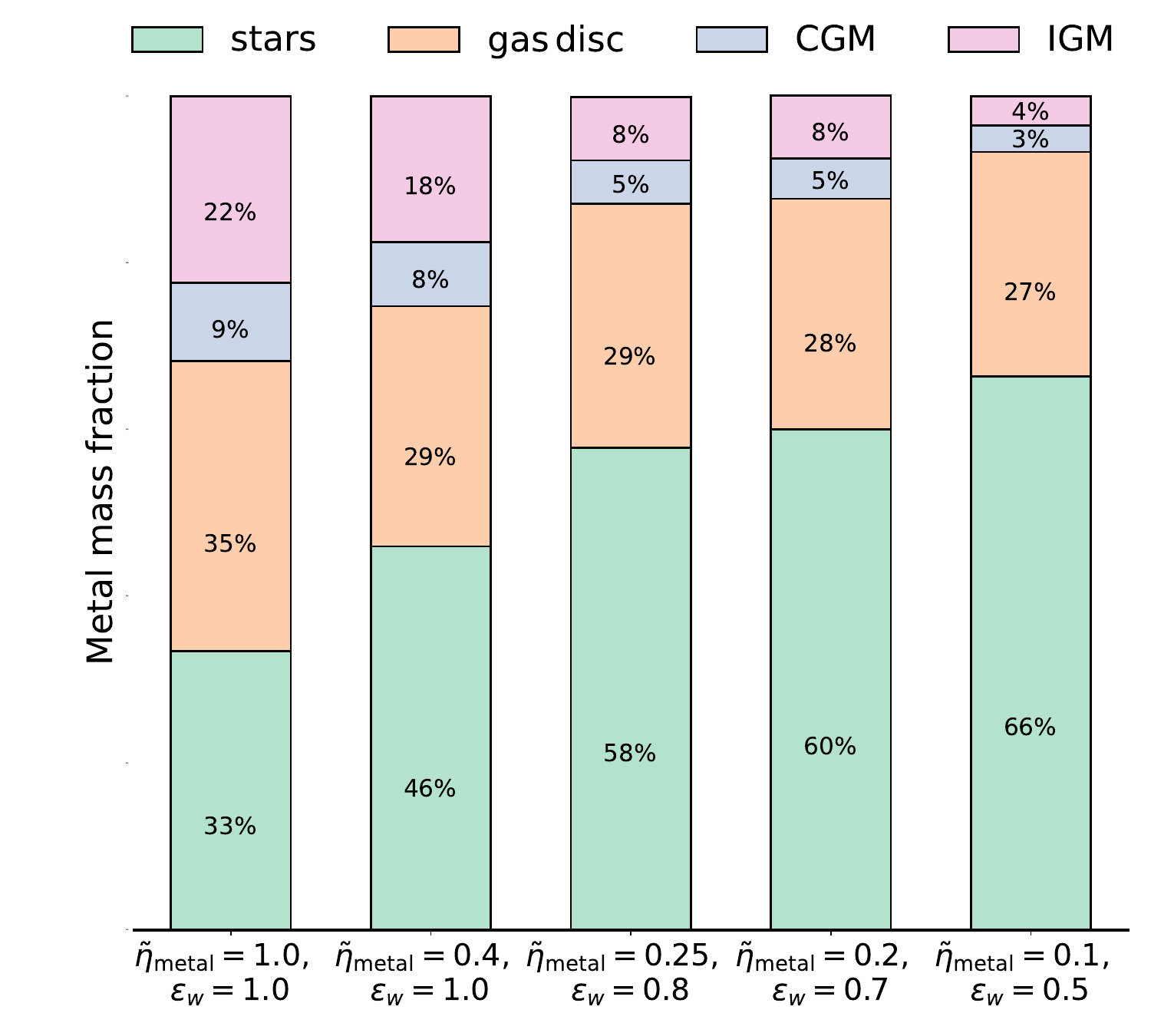}
    \caption{The fraction of total metal mass ($M^{\mathrm{metal}}_{\mathrm{total}}$) ejected by stars over the galaxy's lifetime present in different baryonic components at $z=0$. We split the baryons into four categories: stars (green), gas within the disc (as defined in Table~\ref{tab:definitions}; orange), the CGM (blue) and the IGM (pink), with the percentage of $M^{\mathrm{metal}}_{\mathrm{total}}$ within each component indicated on the respective bars. The $M^{\mathrm{metal}}_{\mathrm{total}}$ of the five runs are reported in Table~\ref{tab:global} and vary by $\lesssim 10\%$ across the runs. From left to right, as the \mlf and \elf decrease, a progressively higher fraction of the metals gets locked up into stars and a smaller fraction reaches the CGM. The fraction of metals escaping the halo into the IGM also drops sharply, showing that IGM enrichment is highly sensitive to the wind parameters. The fraction of metals in the gas disc is nearly constant across our simulations, except for the highest \mlf run on the extreme left. }
    \label{fig:metal_budget}
\end{figure}

\subsection{The metal mass budget}
\label{sec:metal_budget}

\begin{figure*}
	\includegraphics[width=0.95 \textwidth, trim={0 0 0 0}, clip]{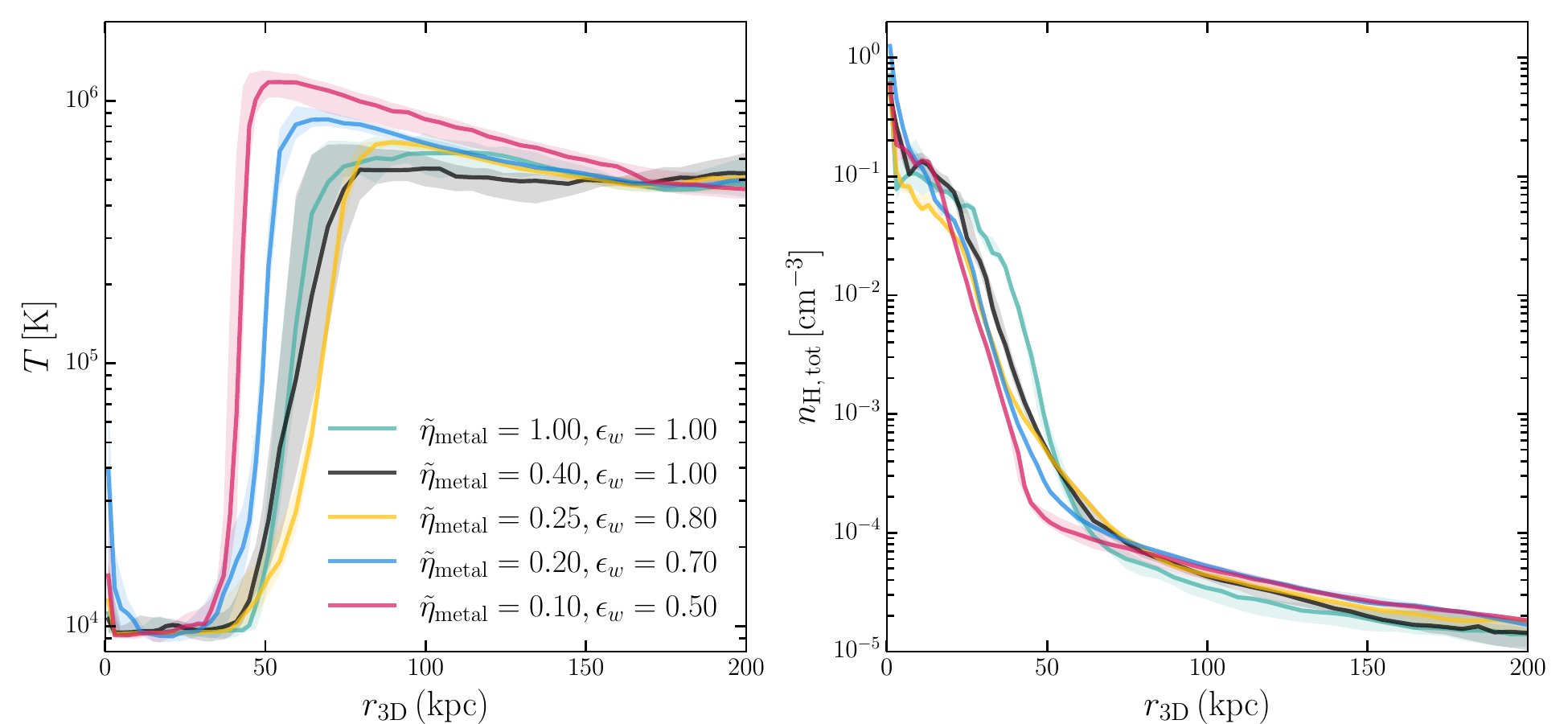}
    \caption{Radial profiles of the gas temperature (left) and hydrogen number density (right) as a function of the galactocentric radius $r_{\mathrm{3D}}$. The profiles are calculated from snapshots spanning the last Gyr of evolution from $z=0.07$ to $z=0$. The solid lines represent the median of the per-snapshot median temperature or number density in a given $r_{\mathrm{3D}}$ bin and the shaded area encloses the 16-84 percentile of the per-snapshot medians. All runs show similar profiles in the inner halo ($r_{\mathrm{3D}} \lesssim 30$ kpc) and also in the outer halo ($r_{\mathrm{3D}} \gtrsim 100$ kpc). In the outer halo, the more metal-poor haloes i.e., those with a lower \mlf exhibit slightly higher temperatures. The temperature profiles feature a sharp jump around 30-50 kpc, but the exact location of this jump depends on \mlf and generally happens at larger radii for higher \mlf runs. The \mlf= 0.25 run does not follow this trend as a result of the consistently lower SFR (by a factor $\sim$ 2) compared to all other runs in the last two Gyr (see Figure~\ref{fig:sfh}). We exclude star-forming gas when computing temperature profiles as its temperature is set by the effective equation of state. At $r_{\mathrm{3D}} \sim 30-50$ kpc, the density profiles show a steep decline. As with the temperature jump, the decline sets in at smaller radii for the lower \mlf runs, reflecting their more compact discs. }  
    \label{fig:temp_density_profile}
\end{figure*}
Galactic winds do not only carry mass, energy, and metals into the CGM but they can also carry these beyond the halo and are responsible for the metal enrichment of the IGM. After inspecting the stellar and gas metallicity profiles across our runs, we now turn our attention to how the global metal budget is distributed across the various baryonic components within the halo and how much of the metals escape into the IGM throughout a galaxy's lifetime. For this, we first calculate the total metal mass, $M^{\mathrm{metal}}_{\mathrm{total}}$, ejected by the stars within each simulated halo over their lifetime as follows:
we identify all star particles within $R_{200c}$ of the galactic centre at $z=0$. For each star particle, the simulation snapshot stores the initial mass, the birth metallicity, and the birth time, from which we can calculate its age at the current time. Using these, we can calculate the total mass of metals that each star particle has ejected into the gas phase over its lifetime by integrating over the IMF and for the three sources of metal enrichment considered in the Auriga model: i) AGB stars; ii) SNII; and iii) SNIa. Summing up the contribution from the three sources gives us $M^{\mathrm{metal}}_{\mathrm{total}}$. 

We find that all our runs produce a similar total metal mass, with $M^{\mathrm{metal}}_{\mathrm{total}} \in 3.1-3.4 \times 10^9$ \msun. This results from their nearly identical final stellar masses and broadly similar star formation histories. We then subtract from $M^{\mathrm{metal}}_{\mathrm{total}}$ the metal mass contained in the gas and stars within the halo at $z=0$. The remaining metals end up enriching the IGM. For this calculation, we consider only stars and gas associated with the main subhalo, excluding contributions from any satellites. This choice only has a marginal effect on our estimates, as the central galaxy dominates both the metal production and the metal reservoir within the halo. Additionally, we split the gas into the disc and CGM components as defined in Table~\ref{tab:definitions}. The total metal mass $M^{\mathrm{metal}}_{\mathrm{total}}$ and the fraction contained in different components are listed in Table~\ref{tab:global} and the metal mass budget at $z=0$ is shown in Figure~\ref{fig:metal_budget}.

Apart from the highest \mlf run, the fraction of metals in the gas disc is nearly constant ($\sim 27-29$ per cent) across our simulations. In contrast, the fraction of metals in stars increases with decreasing metal loading and nearly doubles between the highest and the lowest \mlf runs. The CGM metal fraction instead decreases with decreasing \mlf. 
Despite forming a very similar amount of metals over their lifetime, the five halos vary significantly in their metal retention within the halo at $z=0$. Our highest \mlf run (with \mlf = 1.0, \elf = 1.0) retains 79 per cent of the metals within the halo and allows 22 per cent of the metals to escape into the IGM. This is similar to our fiducial run (\mlf= 0.4, \elf=1.0), which keeps about 82 per cent of the metals within the halo, ejecting the remaining 18 per cent into the IGM. The two intermediate runs with \mlf= 0.25 and \mlf= 0.2 show 92 per cent retention within the halo and exhibit a sharp drop in the IGM enrichment fraction by over a factor of 2 compared to the fiducial run. We further see a factor of 2 drop in the IGM enrichment fraction between these intermediate runs and the \mlf = 0.1 run. The halo in this run retains about 96 per cent of the metals ever produced within the halo
and only ejects 4 per cent into the IGM, a factor of 4 lower than the fiducial run.

We expect that both the metal loading and energy loading of winds regulate the extent of IGM enrichment in our runs. While the metal loading sets how much metals the wind carries, the energy loading determines how far into the CGM, and ultimately the IGM, the wind can propagate. At fixed wind velocity, a reduced energy loading implies a lower mass loading (Table~\ref{tab:runs}), so the winds in our lower \mlf runs carry less momentum than their higher \mlf counterparts, and thus get more easily decelerated by the ambient CGM. Additionally, they encounter more resistance from the relatively more massive CGM  (Table~\ref{tab:global}). 
Both of these effects act to stall the outflow, limiting the escape of metals into the IGM. Thus, we conclude that the IGM enrichment across our runs increases with simultaneously increasing \mlf and \elf.

\begin{figure}
	\includegraphics[width=\columnwidth, trim={0 0 0 0}, clip]{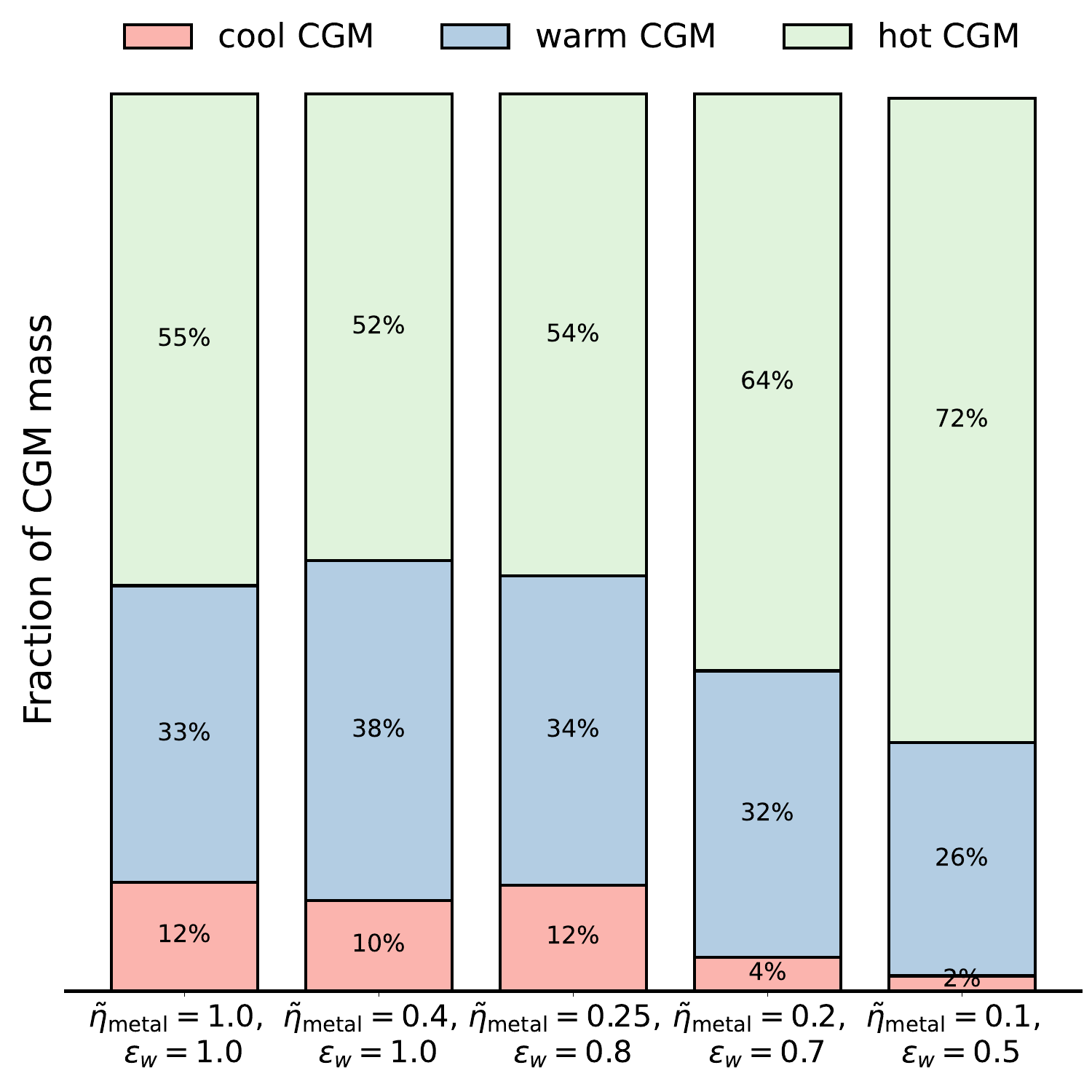}
    \caption{The fraction of CGM gas in different phases averaged over snapshots covering the last Gyr. We split the CGM gas into three phases -- cool CGM (with $T < 5 \times 10^4$ K), warm CGM (with $5 \times 10^4 \, \mathrm{K} \leq T < 5 \times 10^5$ K), and hot CGM (with $T \geq 5 \times 10^5 $K). We exclude all star-forming gas as its temperature is set by the equation of state. We also exclude all gas in satellites. Reducing \mlf below 0.25 leads to a systematic increase (decrease) in the hot (cool) gas fraction of the CGM. }  
    \label{fig:cgm_gas_budget}
\end{figure}

\subsection{The CGM phase structure}
\label{sec:cgm_phase_structure}

Figure ~\ref{fig:temp_density_profile} compares the radial temperature and density profiles of the runs. The profiles are computed using the last eighteen snapshots covering the final Gyr of evolution. Over this time period, the virial radius ($R_{200c}$) of the halos ranges from 207-213 kpc. The shaded region in the figure shows the temporal scatter across these snapshots. In a given snapshot, the gas cells in each radial bin span over an order of magnitude of temperature and density values at most radii (not shown in the figure for clarity).

For the temperature profiles, we exclude star-forming gas because its temperature is set by the effective equation of state. All runs have similar temperatures ($\sim 10^4$ K) in the inner halo set by the cooling function. The runs also feature a sharp jump of about two orders of magnitude within a few kpc around 30-50 kpc. Beyond this, the temperature gradually declines as we move outwards. The exact location of the aforementioned temperature jump varies with \mlf and generally happens at larger radii for higher \mlf runs. The peak temperature attained is also lower for the runs with a higher \mlf, reaching $\sim 10^6$ K in the lowest \mlf run and $\sim 6 \times 10^5$ K in the highest \mlf run. Both trends are consistent with the higher CGM metallicity in the higher \mlf runs, which allows the gas to cool more efficiently out to larger radii (see Figure~\ref{fig:cooling2}). However, these trends are not strictly monotonic: we find that neither the jump location nor the peak temperature in the three highest \mlf runs (\mlf $\geq 0.25$) are ordered by \mlf.

A similar structure is seen in the density profiles that drop sharply across the $30-50$ kpc range. Across this 20 kpc distance, the density falls by more than two orders of magnitude. Beyond this, the decline is more gradual, dropping by about an order of magnitude from $\sim 50$ kpc out to the virial radius ($\sim 200$ kpc). In contrast, the temperature only falls off by a factor of few over the same distance.

These differences in the temperature and density structure are also reflected in the phase composition of the CGM. In Figure~\ref{fig:cgm_gas_budget}, we classify gas as cool at temperatures $T < 5 \times 10^4$ K, warm at $5 \times 10^4$ K $\leq T < 5 \times 10^5$ K, and hot at $T \geq 5 \times 10^5$ K.
We find that the cool gas fraction of the CGM increases with increasing \mlf. This is a direct consequence of the higher metal content of the CGM, which helps cool a large reservoir of CGM gas.

The three highest \mlf runs (\mlf $\geq 0.25$) have broadly similar cool and hot gas fractions despite their different metal and energy loadings. The marginal differences are likely driven by the stochastic SFR in these runs (see Figure~\ref{fig:sfh}). In contrast, further reducing the \mlf has a more pronounced effect on the CGM composition. We speculate that this happens because above a certain CGM metallicity threshold, the cooling is efficient enough and further increasing the CGM metallicity (by increasing \mlf for instance) seems to have no additional impact. This is also reflected in the location of the jump and the peak temperature attained in their temperature profiles, which do not follow a monotonic trend with \mlf for these three runs. We also note that reducing the energy loading of the winds produces a relatively hotter CGM in our runs, which is counter-intuitive. This results from the lower \mlf in these runs which produces a metal-poor CGM with less efficient cooling. We discuss this further in Section~\ref{sec:lit_comparison}.

\subsection{Inflows and outflows}
\label{sec:inflow_outflow}
\begin{figure*}
	\includegraphics[width=\textwidth, trim={0 1cm 0 1.8cm}, clip]{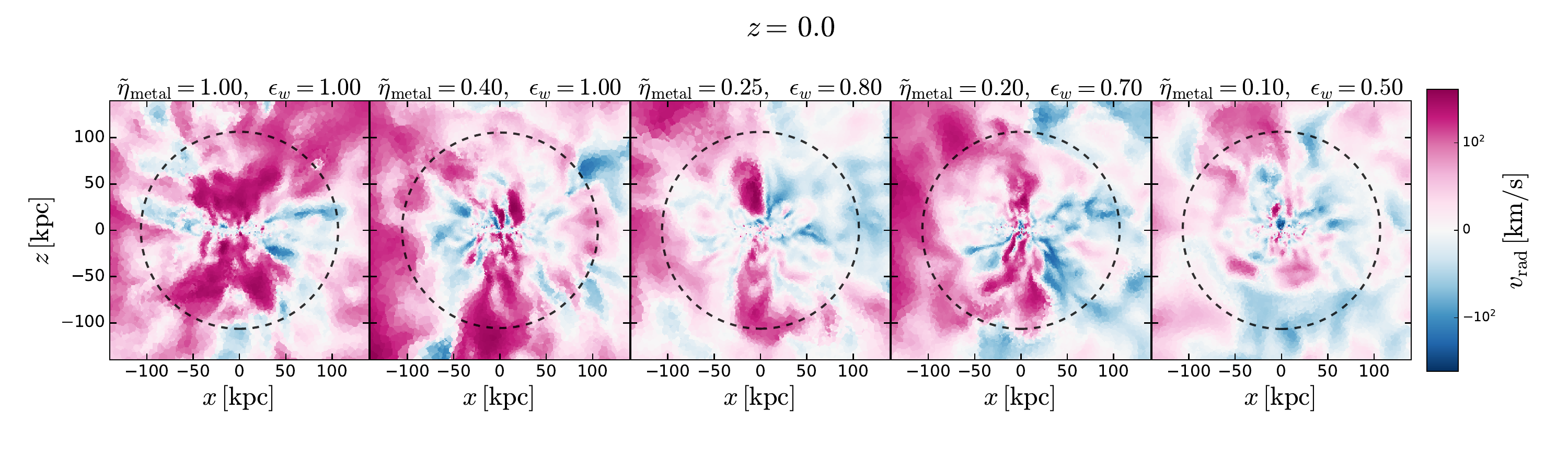}
    \caption{Edge-on radial velocity slices through the centre of the galaxy in the five simulations. The black dashed circle has a radius of $0.5 \, R_{200c}$. The pink and blue colours denote outflowing and inflowing gas, respectively. From left to right, reducing the \mlf and \elf of the winds weakens the outflows as seen from the reduced magnitude of the velocities. Particularly, in the \mlf= 0.1, \elf= 0.5 run, the extent of the outflowing gas below the disc plane is substantially smaller than in the other runs. }  
    \label{fig:vrad_slice}
\end{figure*}
\begin{figure*}
    \includegraphics[width=0.95\textwidth, trim={0 0 0 1.5 cm}, clip]{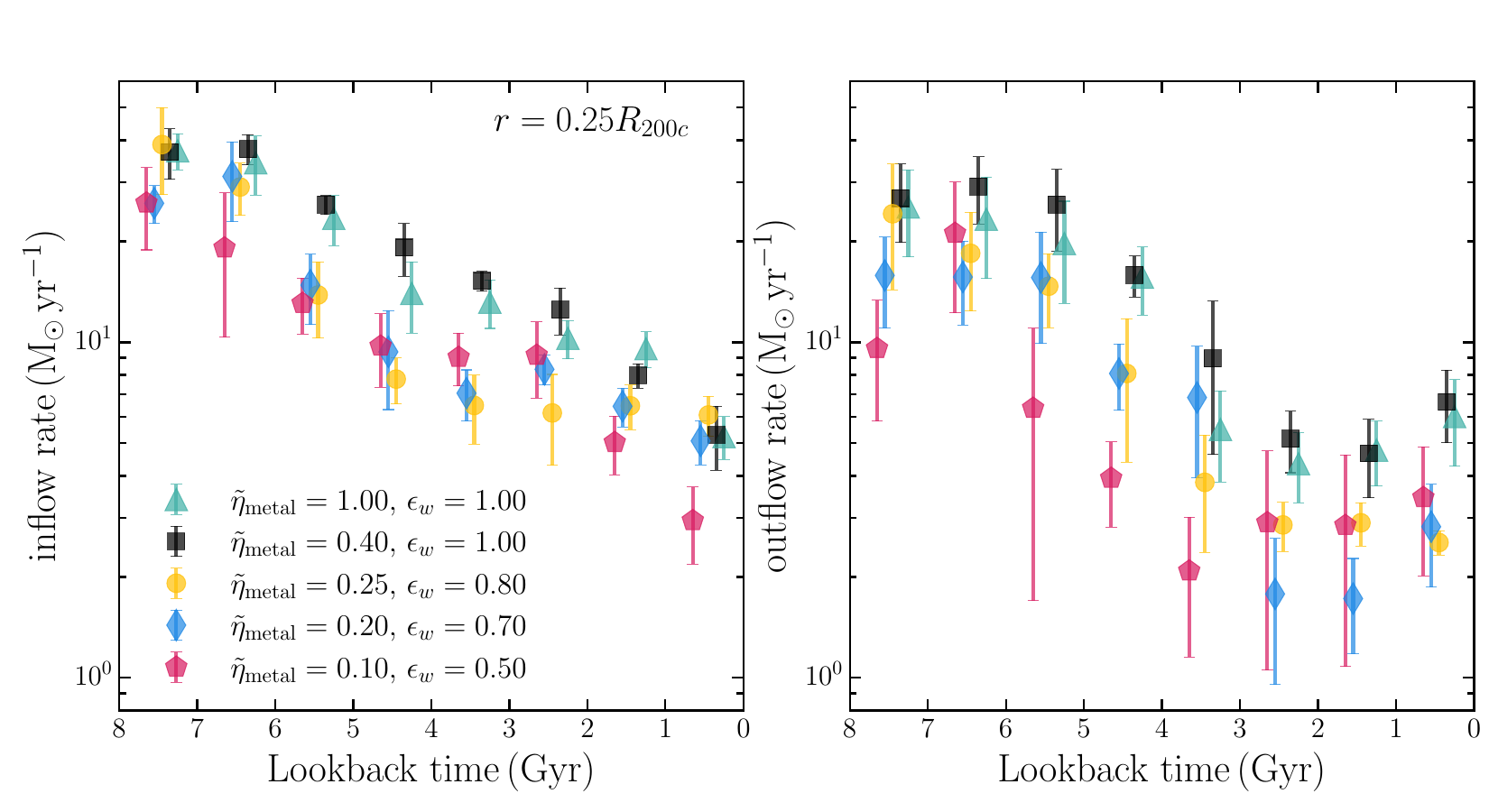}
    \caption{The inflow (left) and outflow (right) rates across a thin spherical shell of width 0.01 $R_{200c}$ located at $r=0.25 \, R_{200c}$ as a function of the lookback time. Each data point shows the flow rate across the shell averaged over snapshots within a 1 Gyr timescale with the bars denoting the $1 \sigma$ standard deviation across these snapshots. The \mlf= 1.0 run and the fiducial run (with identical\elf) show similar inflow and outflow rates at all times while the three lower \mlf runs always has systematically lower inflow and outflow rates compared to the higher \mlf runs.}  
    \label{fig:inflow_outflow_rates}
\end{figure*}

We now turn our attention to the inflow and outflow properties of our galaxies and inspect how the flow of gas across the halo is regulated by the metal and energy loading of the winds. Figure~\ref{fig:vrad_slice} shows edge-on radial velocity slices for the five runs at $z=0$. We find that reducing the metal and energy loading lowers the outflow velocities and the outflowing gas in the lower \mlf runs does not reach as far into the halo. This happens because the lower energy loading of these winds implies a lower mass loading (Table~\ref{tab:runs}). So, although the winds are launched at the same velocity, less mass is ejected per unit star formation. As a result, the outflow gets stalled more easily as it propagated through the ambient halo gas in the lower \elf runs. Additionally, the inflow velocities seem to follow a similar trend, but confirming this requires further investigation beyond the scope of this work.

We further quantify the inflow and outflow rates at a fixed location in the halo ($r = 0.25 \, R_{200c}$). These are computed as follows:
\begin{equation}
    \label{eq:inflow_outflow_rate}
    \dot{M}_{\mathrm{in/out}} = \left| \sum_{\substack{i \in \mathrm{shell} \\ \vec{v}_i \cdot \hat{\hat{r}}_i \lessgtr 0}} m_i \, (\vec{v}_i \cdot \hat{r}_i) \, \frac{S_{\mathrm{shell}}}{V_{\mathrm{shell}}} \right| , \,
\end{equation}
where $i$ denotes the index of gas cells that lie within the spherical shell of width 0.01 $R_{200c}$ centred at $r=0.25 \, R_{200c}$; $m_i$ is the gas mass of the cell, $\vec{v}_i$ is its velocity, and $\hat{r}_i$ is the radial unit vector at the cell centre. $S_{\mathrm{shell}}$ and $V_{\mathrm{shell}}$ are, respectively, the surface area and volume of the thin shell such that $S_{\mathrm{shell}}/V_{\mathrm{shell}} = 1/\Delta r$. Inflow and outflow rates are calculated by summing over cells with $\vec{v}_i \cdot \hat{r}_i <0$ and $\vec{v}_i \cdot \hat{r}_i >0$, respectively. We exclude from the calculation all gas within satellites.

We show in Figure~\ref{fig:inflow_outflow_rates} the inflow and outflow rates across the shell as a function of the lookback time. We plot the mean flow rates averaged over snapshots within 1~Gyr time bins. For all runs, the inflow rates are higher at earlier times and decline (with some fluctuations) to the present day, reflecting the decrease in cosmic gas accretion since $z\sim 1$. 

At any given time, the inflow rates across our runs broadly increase with \mlf. This happens because their more metal-rich CGM allows the gas to cool more efficiently (see Figure~\ref{fig:cooling2}). The resulting cool gas can then accrete onto the galaxy. 
The outflow rates follow a similar trend, increasing with increasing \mlf. Here the driver is the mass loading, rather than the metal content of the winds. Our higher \mlf runs also have a higher \elf and therefore a higher mass loading (Table~\ref{tab:runs}). As a result, they eject more mass per unit star formation than their lower \mlf counterparts, thereby increasing the outflow rates.

While the three lower \mlf runs generally have lower inflow and outflow rates compared to the two higher \mlf runs, the exact ordering of individual runs varies from epoch to epoch.
The most notable exception 
occurs in our \mlf = 0.25, \elf = 0.8 run. At lookback times of 2-5~Gyr, this run exhibits inflow rates consistently lower than all other runs. This sustained suppression significantly reduces its star formation rate in the final two Gyr (see Figure~\ref{fig:sfh}).

The outflow rates show larger temporal fluctuations than the inflow rates in all our runs, as evident from their larger scatter. This is because the two (inflow and outflow) are regulated by physical processes happening on different timescales. Inflow rates are primarily regulated by how fast the CGM gas can cool. This depends on the metal content built over a galaxy's lifetime, through star formation, stellar evolution, and outflows. On the other hand, outflow rates are influenced more strongly by the recent star formation history of the galaxy (Figure~\ref{fig:sfh}).

Finally, the higher inflow rates in our higher \mlf runs are compensated by their correspondingly higher outflow rates, resulting in a similar net flow rate (i.e., inflow $-$ outflow) across all our galaxy halos. Consequently, a similar amount of fuel is available for star formation, consistent with the $\lesssim 10\% $ spread in their stellar masses at $z=0$, despite the substantially different mass and metal loading of the winds in these runs. To summarise, these two parameters influence the gas flows in different ways: the mass loading sets the total mass ejected per unit mass of stars formed (see Equation~\ref{eq:p_sf_wind2}), while the metal loading controls cooling in the CGM and hence the rate of gas accretion. Varying them together changes both the outflow and inflow rates, such that the net flow rate is approximately unchanged.

\section{Discussion}
\label{sec:discussion}
In this work, we examine the impact of simultaneously varying the energy and metal loading of galactic winds on the properties of the CGM. Our findings demonstrate that the energy and metal loading of winds influence the metal distribution and thermal structure of the CGM and regulate the flow of gas in and out of galaxies. This investigation is possible with an explicitly effective galactic wind model. Because we can directly specify the mass, energy, and metal loading of galactic winds, we can vary these independently and in controlled combinations (e.g., to attain similar final stellar masses). In contrast, in models where a galactic wind is an emergent consequence of the energy and/or momentum dumped locally into the ISM \citep[see e.g.,][]{schaye2015, hopkins2018, marinacci2019, schaye2026}, the resulting mass, energy, and metal loading are not specified directly by a model parameter.

\subsection{Implications of our results}
\label{sec:implications}

\begin{figure*}
    \includegraphics[width=0.85\textwidth, trim={0 0 0 1.5 cm}, clip]{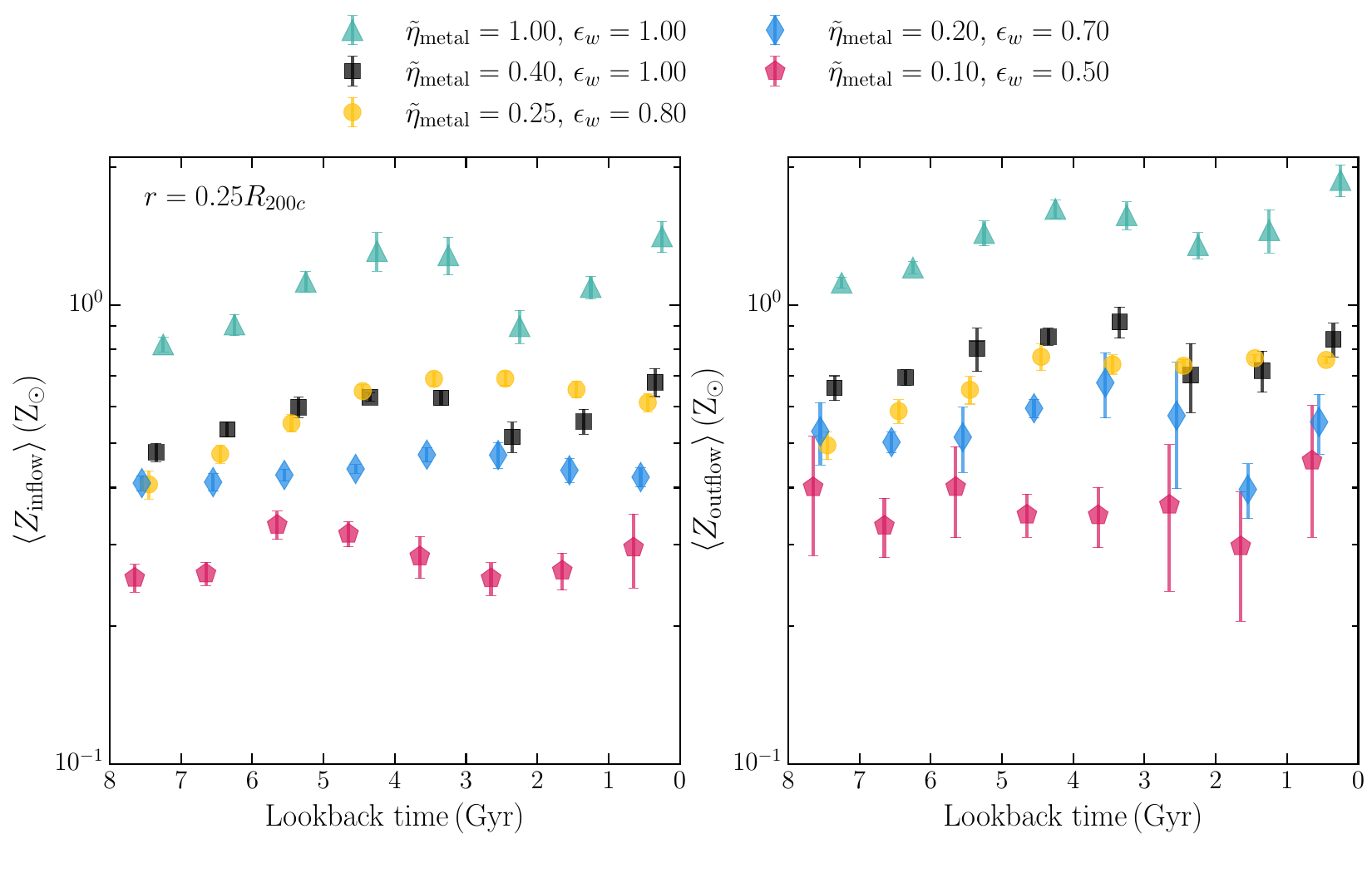}
    \caption{The mean metallicity of inflowing (left) and outflowing (right) gas across a thin spherical shell of width 0.01 $R_{200c}$ located at $r=0.25 \, R_{200c}$ as a function of the lookback time. For each data point, we consider all snapshots within a 1 Gyr bin. The symbols show the mean of the per-snapshot mass-weighted mean metallicities of gas flowing across the shell and the bars denote the $1\sigma$ standard deviation of these per-snapshot means. At any given time, both the inflow and outflow metallicities decrease with decreasing \mlf, except for the \mlf = 0.25 run, which does not follow this trend with respect to the fiducial run. The metallicity of inflowing gas in the two extreme runs are consistently a factor of 4 different at all times.}  
    \label{fig:inflow_outflow_Z}
\end{figure*}
Because our runs have nearly identical final stellar masses, they produce nearly identical total metal masses over their lifetimes (Table~\ref{tab:global}). Despite this, they differ substantially in how the metals are distributed across the baryonic components. Our lower \mlf runs retain more metals is the disc, where they get incorporated into subsequent generations of stars. Consequently, these runs produce more metal-rich stars over time. By $z=0$, the mean metallicity of stars within the halo varies by a factor of $\approx$2 between our two extremes, despite their similar mean stellar ages (Table~\ref{tab:global}). Additionally, the fraction of metals locked up in stars is doubled between the highest and the lowest \mlf runs (Figure~\ref{fig:metal_budget}).

The higher \mlf runs instead expel a higher fraction of metals outside the disc, enriching the CGM and the IGM over time (Figure~\ref{fig:metal_budget}). Figure~\ref{fig:inflow_outflow_Z} shows the mean metallicity of inflowing and outflowing gas across a thin shell at $r=0.25 \, R_{200c}$. The outflow metallicity increases with increasing \mlf. This trend is present partly by construction since \mlf sets the metallicity of the wind. Interestingly, the inflowing gas follows the same order of increasing metallicity with increasing \mlf. This happens because a substantial fraction of the exported metals mix with the ambient halo gas, thereby enriching it. Some of this enriched gas later accretes onto the galaxy.
This pattern is already in place by $z\sim 1$. 

This dependence of the inflow metallicity on \mlf has implications for the interpretation of the fundamental metallicity relation (FMR), where at fixed stellar mass, higher SFR galaxies tend to have lower average ISM metallicities \citep{mannucci2010}. This is typically attributed to the dilution of the ISM metallicity by the pristine or metal-poor gas accreted from the IGM. How metal-poor that gas is by the time it reaches the disc depends on the CGM metallicity. In our suite, the inflow metallicity (at $r = 0.25 \, R_{200c}$) varies by a factor of $\approx$4 between our two extremes. Consequently, galaxies whose winds export metals to the CGM more efficiently would experience less dilution even in the case of an extreme starburst. Thus, we speculate that the extent of the FMR's dependence on the SFR would be influenced by the metal loading of the winds. We note that our runs have been calibrated to have similar star formation histories and nearly identical final stellar masses and therefore, are not suited to test this prediction directly.

The metal retention vs. ejection balance also affects the metallicity gradients within the disc. All our runs show a similar mean metallicity in the extended gas disc ($\sim$1.9-2.2 \zsun) and a similar fraction of the total metals residing there ($\sim$27-29 per cent). The \mlf = 1.0 run is an exception to the this and retains a higher fraction of metals (35 per cent) in the disc simply because it has a more massive disc. Its mean disc metallicity is similar to the other runs, so a higher gas mass translates to a higher metal mass (see Table~\ref{tab:global}). Despite these similarities, our runs differ substantially in their metallicity gradients in the disc. In our highest \mlf run, metals removed from the central parts of the gas disc are later deposited in the outskirts via metal-rich accretion. This lowers the central metallicity while raising it in the outskirts, resulting in a flatter profile. In contrast, our lower \mlf runs experience this redistribution of metals from the disc centre towards the outskirts to a lesser extent. Their outflows are relatively metal-poor, their accretion rates are lower, and the gas deposited at larger radii is correspondingly less enriched. Moreover, because of the relatively lower wind recycling experienced by these runs, their discs are not as extended as in the highest \mlf run, which further steepens the gradients. Overall, their profiles are therefore more centrally concentrated, with steeper gradients.

The flattening of metallicity gradients via metal-rich outflows has been reported in previous studies as well. Using a semi-analytical model, \cite{fu2013} showed that the metals ejected into the halo enrich the gas that later accretes onto the disc, causing a flattening of the present-day gradients. \cite{acharyya2025} obtain similar results using the FOGGIE simulations.
We note that flat metallicity gradients can also arise from gas-rich mergers through nuclear dilution by metal-poor inflows \citep{rupke2010, montuori2010, bustamante2018}. While this is not the case for our runs, observationally, a measured gradient alone would not distinguish the two scenarios.

\subsection{Comparison with observations}
\begin{figure}
	\includegraphics[width=0.95\columnwidth]{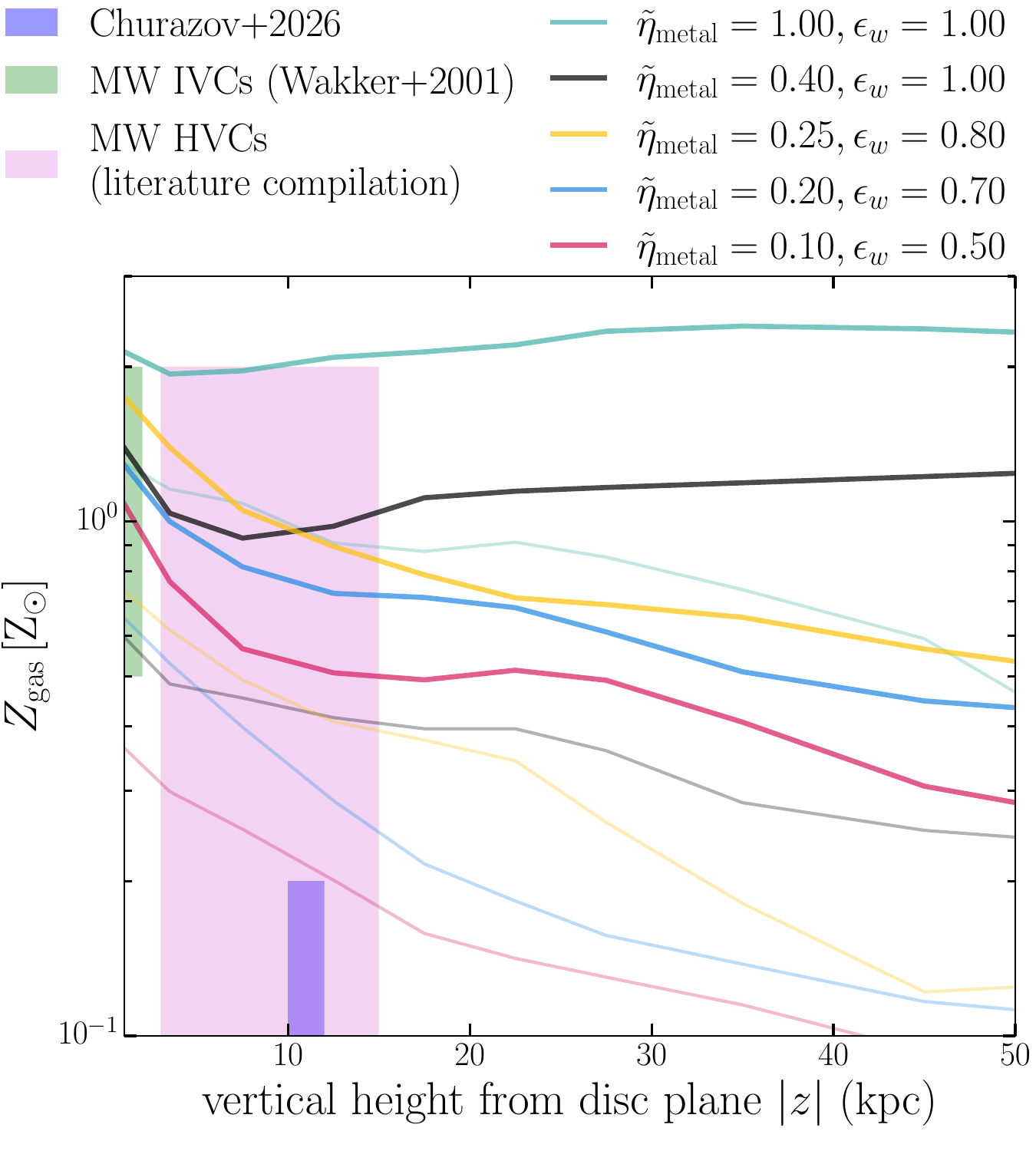}
    \caption{Gas metallicity (\Zgas) profiles as a function of the vertical distance $|z|$ from the disc plane for gas within $r_{\mathrm{2D}} \leq 50$ kpc of the galactic centre. As before, the solid line shows the median of the per-snapshot medians in a given $|z|$ bin. The thin coloured lines show the median of the per-snapshot 2.5th percentiles, indicating the lower-metallicity end of the gas at each $|z|$. The three lower \mlf runs decline steadily with increasing $|z|$, whereas the two higher \mlf runs remain fairly flat. As a result, at $|z| \sim 50$ kpc, there is almost an order of magnitude difference between the median metallicity of our two extreme runs (\mlf = 1.0, \elf = 1.0 and \mlf  = 0.1, \elf = 0.5). The observational constraints shown in coloured rectangles come from measurements based on X-ray observations of the south-eastern eROSITA bubble \citep{churazov2026} and from absorption-line measurements of intermediate- and high-velocity clouds in the Milky Way \citep[from][and a compilation of studies listed in the main text]{wakker2001}. Since neither the IVCs nor the HVCs are known to be forming stars, we exclude star-forming gas from this analysis.}
    \label{fig:gas_metallicity_profile_z}
\end{figure}

The azimuthal variation in the CGM metallicity reported in Figure~\ref{fig:gas_metallicity_maps} translates directly into differences among the vertical metallicity profiles of our runs. Figure~\ref{fig:gas_metallicity_profile_z} shows the \Zgas profiles as a function of the vertical distance above the disc plane, $|z|$, for gas within $r_{\mathrm{2D}} \leq 50$ kpc of the galactic centre. Our \mlf = 1.0 run maintains a nearly flat profile out to $|z| = 50$ kpc, with a median metallicity of $\approx$2 \zsun. In contrast, in our \mlf = 0.1 run, \Zgas declines steeply from solar values near the midplane to $\lesssim 0.5$ \zsun by $|z| = 20$ kpc. By $|z| \sim 50$ kpc, the two extremes differ by almost an order of magnitude. This results from the lower metallicity of the outflowing gas, combined with a narrower opening angle of the metal-rich component.

Existing measurements of the metallicity of the Milky Way halo gas are sparse and largely confined to small $|z|$. Intermediate- and high-velocity clouds (IVCs and HVCs) are identified in absorption against background QSOs or halo stars, and their metallicities are derived by comparing metal ion column densities against neutral hydrogen, with ionisation correction and photoionisation modelling. IVCs, which lie within $|z| \lesssim 1.5$ kpc \citep{lehner2022}, are found to have near-solar metallicities \citep{wakker2001}. HVCs are found farther out, at $|z| \sim 5-15$ kpc \citep{lehner2022}, and span a broader range of metallicities, from a few percent solar to supersolar \citep{wakker2001, richter2001, zech2008, yao2011, tripp2012, fox2016}. The median metallicities of our runs show comparable values at these heights, although reaching only the higher part of the HVC metallicity range. However, as the gas in the CGM is far from homogeneous, we find a large spread in the metallicity of gas cells at a given height. Thus, the 2.5th percentiles of our lower \mlf runs extend towards the lower part of the observed HVC metallicity range. These are shown as thin lines in the figure.

At $|z| \sim 10-12$ kpc, \cite{churazov2026} model the south-eastern part of the eROSITA bubbles as a forward shock propagating through the CGM and constrain the metallicity of the shocked gas to $\lesssim 0.1$ \zsun, with a systematic uncertainty of a factor of 2. Because it is a local rather than a line-of-sight measurement, this provides an in-situ constraint on the Milky Way CGM at these heights. At this distance, the median metallicity in our runs ranges from $\sim$0.5 \zsun in the \mlf = 0.1 run to supersolar values in the \mlf = 1.0 run. The 2.5th percentile in our \mlf= 0.1 run reaches values comparable to \cite{churazov2026}, implying that only about 2 per cent of the gas by mass in our lowest \mlf run at these heights is as metal-poor as these observations suggest.

A one-to-one comparison with either set of measurements is not straightforward. The IVC and HVC metallicities trace individual clouds along single sightlines, selected by velocity, whereas our profiles are mass-weighted medians of all halo gas at a given height. Likewise, the eROSITA constraint probes gas in a transient shocked state. Therefore, a fair comparison would require identifying discrete cloud-like structures in our simulations and generating synthetic absorption-line measurements, which we defer to future work. 

The current constraints therefore neither favour nor rule out any of our models. At $|z| \lesssim 15$ kpc, where measurements exist for several IVCs and HVCs, the median \Zgas in our models varies by a factor of $\sim$4, comparable to the spread among the individual clouds themselves. Our models differ more strongly at larger $|z|$ ($\gtrsim 30$ kpc), where the three lower \mlf runs decline steadily with height, while the two higher \mlf runs show nearly flat profiles out to $|z|\sim 50$ kpc. Metallicity measurements of HVCs at these heights would be useful to distinguish between our wind feedback models. In addition, it would be helpful to obtain more in-situ constraints such as those derived by \cite{churazov2026}, which at present is based on a single measurement.


\subsection{Comparison with other simulations}
\label{sec:lit_comparison}
We now situate our findings in the context of previous numerical work that investigated whether the CGM retains imprints of galactic feedback.

\cite{suresh2015} investigate the CGM properties of the galaxies in a 25~$h^{-1}$~Mpc cosmological box, using different feedback variants, including changes to the metal loading of galactic winds. 
Similar to us, they model galactic winds using wind particles. Their pristine winds model (\mlf = 0.0) produces a factor of $\approx$5 lower gas metallicity in the inner halo than their fully-enriched winds model (\mlf=1.0), along with less cool gas in the CGM. We find the same trend: a factor of $\approx$4 difference in metallicity between our highest and lowest \mlf runs at similar radii (Figure~\ref{fig:gas_metallicity_profile}) as well as a significantly lower cool gas fraction in our \mlf = 0.1 run (Figure~\ref{fig:cgm_gas_budget}).

Their simulation setup is comparable to ours. They use \textsc{arepo} and the same wind feedback prescription from \cite{vogelsberger2013} that the Auriga model derives from. However, the key difference is that they do not keep the stellar mass fixed across their runs. They find that the metal loading of winds impacts the cosmic star formation rate density by altering the efficiency of metal-line cooling and hence, the strength of the galactic fountain in their galaxies at late times. We find that a similar mechanism regulates the inflow rates and the size of the extended gas disc in our runs. By varying the metal loading in their suite, they also vary the total metal mass produced across their runs. Instead, by changing the metal loading and energy loading of winds in tandem to obtain a fixed $z=0$ stellar mass (within $\sim$10\%),  we hold the total metal mass produced fixed. The differences we obtain in the CGM properties of our galaxies therefore arise at fixed total metal production.

Using the ARKENSTONE framework \citep{smith2024} for modelling a hot wind, \cite{bennett2025} independently vary the energy loading and mass loading of galactic winds to compare ejective (low specific energy winds) and preventive (high specific energy winds) feedback in a cosmological box of side length 39.6 Mpc. They find that their higher specific energy winds heat and deplete the CGM, suppressing further accretion, and thereby regulating the star formation. Here as well they do not keep the stellar mass fixed across their runs and find that the cosmic stellar masses of their galaxies at fixed halo mass are sensitive to the energy loading of winds.
Unlike \cite{bennett2025}, we hold the specific energy of winds fixed across our suite (since we vary the \elf at fixed wind velocity; see Table~\ref{tab:runs}), and find that the metal composition of the CGM, rather than the wind energetics, drives the temperature structure and accretion rates in the CGM. The more metal-poor CGM of our lower \mlf runs leads to longer cooling timescales (Figure~\ref{fig:cooling2}), which slows down accretion onto the disc. We cannot test the impact on star formation as we hold the stellar mass nearly identical by construction.

\cite{rey2025} simulate a $5 \times 10^{11}$ \msun halo to $z=1$ with three supernova feedback prescriptions -- purely mechanical, purely thermal (with delayed cooling), and a hybrid of the two. Their runs are calibrated to yield the same stellar masses at $z=1$, though with a spread of 50 per cent between their extreme cases, compared with 10 per cent across our runs. Similar to us, they find that galaxies matched in stellar mass exhibit significantly different CGM properties, including the inflow and outflow rates and the phase structure of the halo gas. Likewise, their metal budgets show a comparable spread to our Figure~\ref{fig:metal_budget}. Their mechanical and hybrid models eject only 5-10 per cent into the IGM and retain $\gtrsim 70$ per cent in stars, similar to our lower \mlf runs (\mlf $\leq 0.25$), which also allow $\lesssim 10$ per cent of the metals to escape to the IGM. Their delayed cooling model instead expels 64\% of the metals into the IGM, while retaining only $\approx$18 per cent in stars by $z=1$. This model drives stronger metal outflows than even our most extreme (\mlf =1.0) run, though the authors caution that this prescription is overly efficient and a more realistic model would retain a higher fraction of metals within the halo. 
Overall, our findings are qualitatively consistent with them despite the two studies varying different aspects of stellar feedback modelling. They contrast different feedback mechanisms, whereas we hold the mechanism fixed, and vary only the energy and metal content carried by the winds. 

\vspace{0.5cm}
To summarise, together these studies and ours reaffirm that feedback models that produce indistinguishable stellar masses can vary substantially in their CGM properties. Thus, the CGM retains a signature of the feedback physics within cosmological simulations. 

\subsection{Physical origin of a low-metallicity wind}
\label{sec:low_Z_wind}
Our effective wind model is agnostic to what drives the wind and our different runs could, in principle, correspond to different driving sources. While it is beyond the scope of this work to identify these sources, it is interesting to consider how a low metal loading wind could arise physically. One possibility is that the wind material is not launched from the disc midplane but rather from a few kpc above the plane. The ambient gas there is a mixture of the enriched outflows and low-metallicity halo gas and as such would be less enriched than the midplane ISM. If the wind is driven from this location, i.e., it is not direct SN ejecta, it could have a lower metallicity. 

Using simulations of an isolated Milky Way-mass galaxy that resolve the injection and propagation of thermal and cosmic-ray feedback, \cite{thomas2025} find that winds are accelerated $\sim 2-3$ kpc above the disc plane. In their simulations, a cosmic-ray-driven wind entrains substantially more gas from the inner CGM than a thermal wind. Therefore, a wind driven largely by cosmic rays could have a considerably lower metallicity compared to the wind driven from the midplane ISM instead. 

In addition, high-resolution simulations that resolve the multiphase ISM and inject feedback locally \citep[e.g.,][]{kim2017, rathjen2023, vijayan2026} 
find that the resulting outflows are themselves multiphase. These studies show that most of the outflowing mass is carried by the cool component while the newly-synthesised metals reside in the hot phase \citep{kim2020}. The mass-dominating cool component of the outflow would therefore have a lower metal loading on average than the hot phase. Likewise, \cite{vijayan2026} find that the metal loading depends on the nature of the outflow. The cool and bursty outflows in their simulations have a factor of 4-5 lower metal loading compared to sustained, hot or multiphase outflows. They attribute this to supernovae exploding in denser environments in the former case, where radiative losses prevent the hot gas from escaping the disc. Thus, the escaping outflow is both cooler and more metal-poor. 

Together these studies support the plausibility of low-metallicity winds, but do not confirm the metallicity values we attain. This would require a direct comparison of the outflow metallicity in those simulations against ours. More importantly, this would need to be tested in a fully cosmological setup where the wind is launched into and propagates through a multi-phase, evolving CGM, rather than the idealised or isolated environments these studies employ. We leave this comparison to future work.

\section{Conclusions}
\label{sec:conclusions}
In this work, we have demonstrated that more than one set of wind feedback parameters (\mlf, \elf) in the Auriga galaxy formation model can reproduce the stellar component of a Milky Way-like galaxy. Our galaxies have similar morphologies and stellar masses agreeing within $\approx$10\% and therefore produce nearly identical total metal mass throughout their lifetimes. Our simulations differ substantially in how these metals are distributed among the baryonic components (stars, gas disc, CGM, and IGM) and throughout the gaseous halo. 
These differences arise from the different wind parameters. The metal loading of winds sets the balance between metals retained within the disc and those ejected into the CGM, while the energy loading governs how far the winds propagate. Together they regulate 
how much metals escape the halo and end up enriching the IGM.  
Our key findings are as follows: 
\begin{enumerate}
    \item The runs show stark differences in the gas-phase metal distribution, both within the extended gas disc (Figure~\ref{fig:gas_metallicity_profile}) and as a function of the height above the disc plane (Figure~\ref{fig:gas_metallicity_profile_z}, see also Figure~\ref{fig:gas_metallicity_maps}). The extended discs have different metallicity gradients, with the lower \mlf runs showing more centrally concentrated metallicity profiles, with a steeper decline.
    \item Simultaneously decreasing the metal loading and energy loading of winds leads to progressively more metals being locked up in stars. A lower amount of metals are exported out of the disc, resulting in a less enriched CGM and IGM over time (Figure~\ref{fig:metal_budget}). Our lowest metal loading run (\mlf = 0.1) contains $\approx$66 per cent of the metals locked up in stars by $z=0$, compared to only $\approx$33 per cent for our highest metal loading (\mlf = 1.0) run. The former allows only 4 per cent of the metals to leave the halo over the lifetime, while the latter allows for 22 per cent of the metals ever produced to reach the IGM.
    \item Increasing the metal loading of winds increases the cooling efficiency of the CGM gas, resulting in higher accretion rates onto the galaxy as well as higher metallicity of the accreting gas (see Figures~\ref{fig:inflow_outflow_rates} and \ref{fig:inflow_outflow_Z}). Combined with the higher mass loading of these runs (Table~\ref{tab:runs}), which drives more gas through the galactic fountain, this produces more extended gas discs in our higher \mlf runs.
    \item Varying  the wind parameters alters the phase structure of the CGM. For \mlf $\leq 0.2$, there is a significant decline of the cool gas fraction within the CGM, accompanied by a rise in the hot gas fraction (Figure~\ref{fig:cgm_gas_budget}). Contrary to our expectation, reducing the energy loading  and metal loading of the winds at the same time results in a hotter CGM.
    
\end{enumerate}
The CGM holds an imprint of the feedback processes taking place within galaxies. We expect these differences in the CGM properties and metal distribution to translate into differences in observables such as the column densities of metal ions tracing different CGM phases. Future work on predicting such observables diagnostics from simulations would allow us to investigate whether these are able to distinguish between wind feedback models. On the observations side, measurements of the Milky Way halo gas metallicity at vertical heights $|z|\gtrsim 30$ kpc, where our runs diverge most strongly, would be useful for constraining the metal loading of galactic winds. The CGM therefore provides a promising independent avenue for distinguishing between feedback models that remain degenerate when considering the stellar properties of galaxies alone.

\section*{Acknowledgements}
PK and FvdV are supported by a Royal Society University Research Fellowship (URF\textbackslash R\textbackslash241005). 
RB is supported by the SNSF through the Ambizione Grant PZ00P2\_223532. RJJG acknowledges support from an STFC Ernest Rutherford Fellowship (ST$/$W003643$/$1): `GalaHAD: Galaxy formation with High Accuracy Dynamics'. 
TAR is supported by the UKRI CDT in Artificial Intelligence, Machine Learning and Advanced Computing (\href{https://cdt-aimlac.org/}{AIMLAC}), funded by grant EP/S023992/1. 
This work used the DiRAC@Durham facility (under project code dp424) managed by the Institute for Computational Cosmology on behalf of the STFC DiRAC HPC Facility (\hyperlink{www.dirac.ac.uk}{www.dirac.ac.uk}). The equipment was funded by BEIS capital funding via STFC capital grants ST/P002293/1, ST/R002371/1 and ST/S002502/1, Durham University and STFC operations grant ST/R000832/1. DiRAC is part of the National e-Infrastructure.

\section*{Data Availability}
The data underlying this article is owned by the SURGE collaboration and will be shared upon reasonable request to the corresponding author. 



\bibliographystyle{mnras}
\bibliography{example} 




\appendix
\label{sec:appendix}

\section{Gas surface density}
\begin{figure*}
	\includegraphics[width=\textwidth, trim={0 1cm 0 0}, clip]{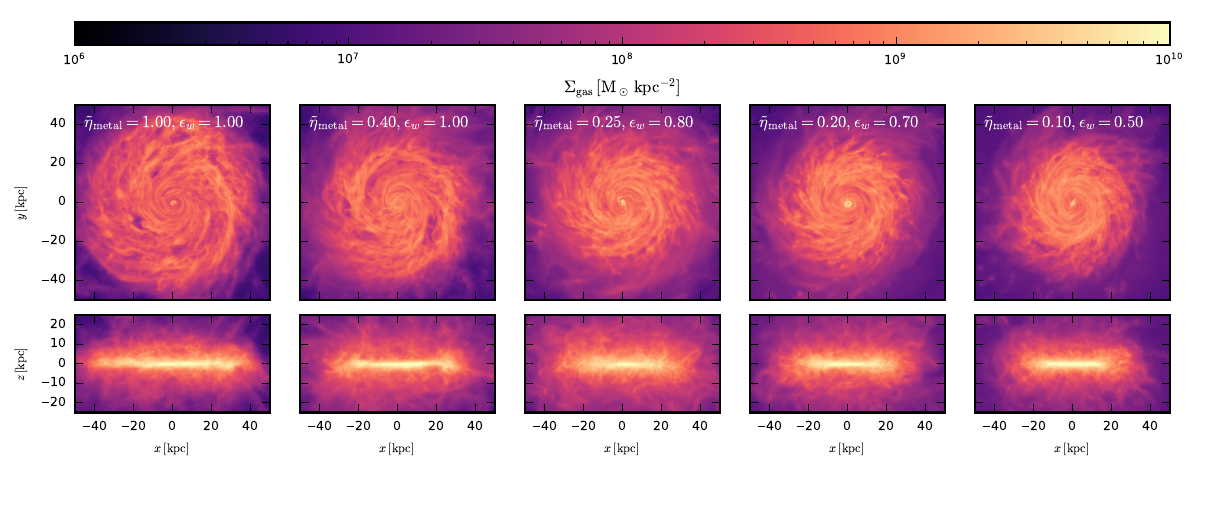}
    \caption{Face-on (top) and edge-on (bottom) projections of the gas mass surface density for the different runs. The run with the highest \mlf on the extreme left produces a substantially larger gas disc compared to the other runs. The rightmost run with the lowest \mlf forms a much more compact disc. The gas masses within a disc of radius 50 kpc and height 10 kpc above and below the disc plane are reported in Table~\ref{tab:global}.}
    \label{fig:gas_maps}
\end{figure*}

Figure~\ref{fig:gas_maps} shows the face-on and edge-on views of the gas mass surface density maps of the five galaxies in our suite. From left to right, both the metal loading parameter \mlf and the energy loading parameter \elf of winds decrease (except for the first two runs which have the same \elf). We find that the higher \mlf runs feature more extended gas discs, similar to their more extended stellar discs (Figure~\ref{fig:stellar_maps}), while the gas discs for the lower \mlf runs are more compact. The higher metal content of the CGM in the higher \mlf runs aids the cooling of gas, which then accretes onto the disc and builds up its mass over time. This also results in more gas cycling out of and back into the disc. In the Auriga model, such a fountain flow enhances the angular momentum of the gas \citep{grand2019}. This gas is deposited on the disc outskirts leading to a more extended gas disc over time.

\section{Cooling time of CGM gas}
\begin{figure*}
    \centering
	\includegraphics[width=\textwidth, trim={0 0 0 0}, clip]{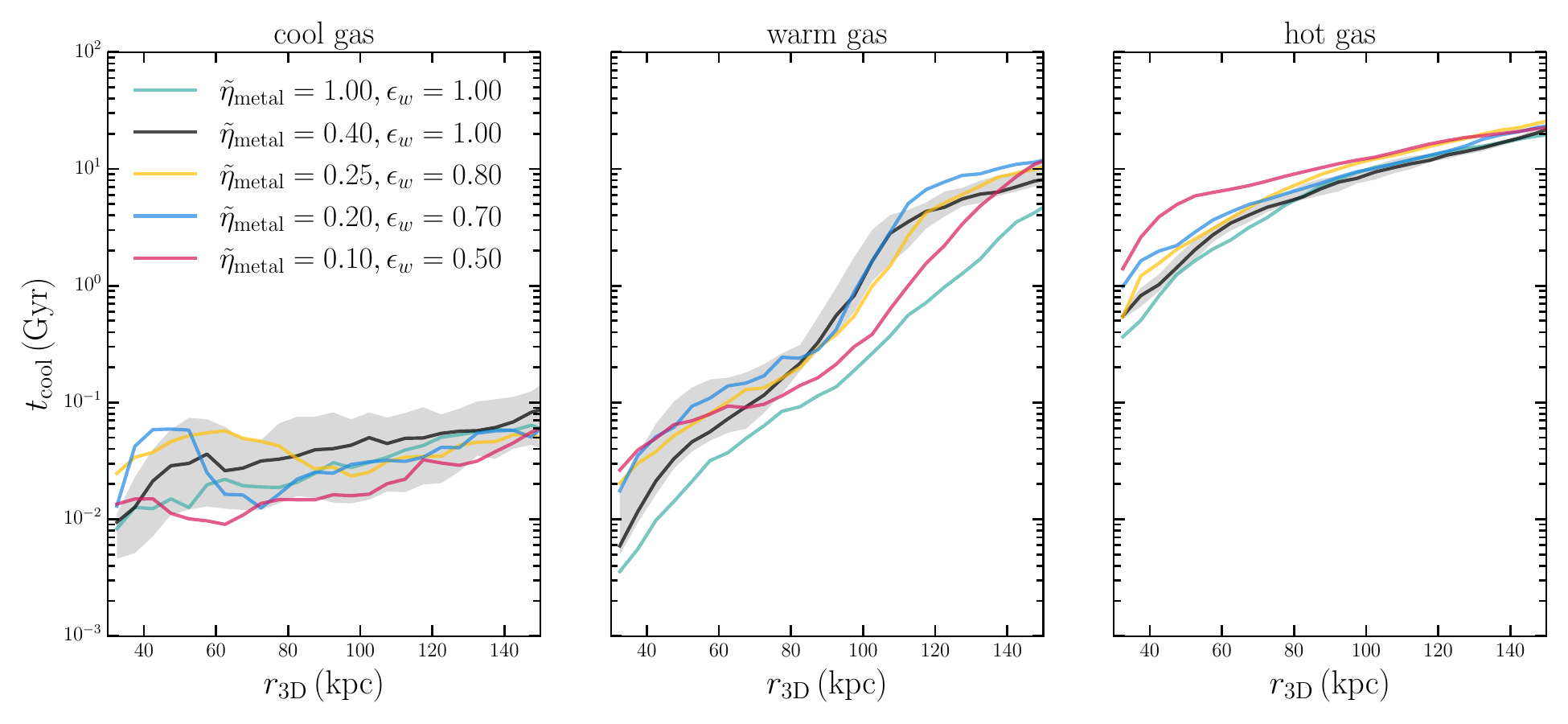}
    \caption{The cooling time for different gas phases (cool: $T<5 \times 10^4$ K, warm: $5 \times 10^4$ K $\leq T<5 \times 10^5$ K, hot: $T\geq 5 \times 10^5$ K) as a function of the 3D galactocentric radius, $r_{\mathrm{3D}}$, at $z=0$. We use the last 18 snapshots covering the final Gyr of evolution for each run. For each snapshot, we compute the mass-weighted median $t_{\rm cool}$ in each radial bin, and then compute the (unweighted) median of the per-snapshot medians. This is shown as a solid line. The grey shaded region denotes the temporal scatter i.e., the 16-84 percentile of the per-snapshot medians for the fiducial run. Other runs show a similar level of scatter. The cooling time for cool gas is similar across runs. In the inner halo ($r_{\mathrm{3D}} \lesssim 60$ kpc), the cooling time for warm gas tends to increase with decreasing \mlf. For hot gas, the cooling time increases with decreasing \mlf at all radii. This follows from the lower metal content of the CGM in the lower \mlf runs, which significantly reduces the cooling efficiency. This in turn affects the accretion rates onto the galaxy in these runs as shown in Figure~\ref{fig:inflow_outflow_rates}.
    }
    \label{fig:cooling2}
\end{figure*}

Figure~\ref{fig:cooling2} shows the median cooling time, $t_{\mathrm{cool}}$, as a function of the 3D galactocentric radius, $r_{\mathrm{3D}}$. In each radial bin, we split the gas into cool ($T<5 \times 10^4$ K), warm ($5 \times 10^4$ K $\leq T<5 \times 10^5$ K), and hot ($T\geq 5 \times 10^5$ K) phases. For each phase, we show the median cooling time over the last 18 snapshots covering the final Gyr of evolution. The solid lines show the median of the per-snapshot medians while the shaded band (only shown here for the fiducial run) denotes the 16-84 percentile of these medians, reflecting the temporal scatter over this time period. The cooling time of cool gas is similar across runs at all radii with no systematic trends with \mlf and \elf. This is because gas at these temperatures mostly cools via hydrogen and helium cooling lines.

Warm gas tends to show some trends in the inner halo ($r_{\mathrm{3D}} \lesssim 60$ kpc), where the three lower \mlf runs exhibit a longer cooling time compared to the two higher \mlf runs. Beyond that the ordering does not hold and our two extremes show lower cooling times than the other three runs.

The hot gas shows monotonically decreasing $t_{\mathrm{cool}}$ with increasing \mlf at all radii. Overall, the higher cooling efficiency of the halo gas in the higher \mlf runs results in their higher accretion rates seen in Figure~\ref{fig:inflow_outflow_rates}.

For the higher \mlf runs, the higher metallicity of the wind carries more metal enriched gas outside the galactic disc. This metal-rich gas mixes with the ambient CGM gas and enhances the CGM metallicity (see Figures~\ref{fig:gas_metallicity_profile} and \ref{fig:gas_metallicity_profile_z}). The higher metal content enhances metal-line cooling in these runs and consequently reduces the cooling time. We use a relatively wide range of temperatures in our definition of warm gas, and thus the median $t_{\mathrm{cool}}$ of such gas likely depends more strongly on how it is distributed with respect to the peaks in the cooling curve. As a result, $t_{\mathrm{cool}}$ increases with decreasing \mlf only for a narrow range in the inner halo and not throughout the halo. 

%


\bsp	
\label{lastpage}
\end{document}